\documentclass[twocolumn,aps,prc,superscriptaddress,showpacs,floatfix,longbibliography,nofootinbib]{revtex4-1}
\usepackage{url}
\usepackage{cancel}
\usepackage[colorlinks,linkcolor=blue,citecolor=blue,filecolor=black,urlcolor=blue]{hyperref}
\usepackage{epsfig,graphics}
\usepackage{graphicx}% Include figure files
\usepackage{dcolumn}% Align table columns on decimal point
\usepackage{bm}% bold math
\usepackage[usenames]{color}
\usepackage{amssymb}
\usepackage{amsmath}
\usepackage{multirow}
\usepackage{float}
\usepackage{harpoon}
\usepackage{MnSymbol}
\usepackage{appendix}
\usepackage{color}
\usepackage{hyperref}
\usepackage{cleveref}

\begin{document}

\title{Investigating the possibility of extracting neutron-skin thickness in nuclei by their collisions at intermediate energies}
\author{Tian-Ze Li}
\affiliation{School of Physics Science and Engineering, Tongji University, Shanghai 200092, China}
\author{Lu-Meng Liu}
\affiliation{Physics Department and Center for Particle Physics and Field Theory, Fudan University, Shanghai 200438, China}
\author{Jun Xu}\email[Correspond to\ ]{junxu@tongji.edu.cn}
\affiliation{School of Physics Science and Engineering, Tongji University, Shanghai 200092, China}
%\affiliation{Shanghai Institute of Applied Physics, Chinese Academy of Sciences, Shanghai 201800, China}
\author{Zhong-Zhou Ren}%\email[Correspond to\ ]{zren@tongji.edu.cn}
\affiliation{School of Physics Science and Engineering, Tongji University, Shanghai 200092, China}
%\affiliation{Key Laboratory of Advanced Micro-Structure Materials, Ministry of Education, Shanghai 200092, China}
\date{\today}
\begin{abstract}
Inspired by various studies on extracting the density distributions of nuclei from their collisions at ultrarelativistic energies, in the present work we investigate the possibility of extracting the neutron-skin thickness $\Delta r_{np}$ in nuclei by their collisions at intermediate energies. We have analyzed the free neutron-to-proton yield ratio $n/p$ as a candidate probe at both midrapidities and forward rapidities in peripheral and central $^{124}$Sn+$^{124}$Sn collisions based on an isospin-dependent Boltzmann-Uehling-Uhlenbeck (IBUU) transport model, and found that the resulting $n/p$ yield ratio is more sensitive to the symmetry potential in the collision dynamics than to the initial $\Delta r_{np}$ in colliding nuclei in most cases. The largest effect on the $n/p$ yield ratio from the initial $\Delta r_{np}$ is observed for nucleons at large transverse or longitudinal momenta in central collisions at the collision energy of a few GeV/nucleon.
\end{abstract}
\maketitle

\section{introduction}

Extracting the nuclear matter equation of state (EOS) has been a fundamental purpose of the nuclear physics for decades. The nuclear symmetry energy $E_{sym}$, which characterizes the difference in the binding energy per nucleon between isospin symmetric and asymmetric matter, remains the most uncertain part of the nuclear matter EOS, while it has important ramifications in nuclear structures, heavy-ion reactions, and nuclear astrophysics~\cite{Li:2014oda,Steiner:2004fi,Baran:2004ih,Li:2008gp,Baldo:2016jhp}. The neutron-skin thickness $\Delta r_{\mathrm{np}}$, defined as the difference between the neutron and proton root-mean-square (RMS) radii, is one of the most robust probes in constraining the slope parameter $L=3\rho_0 (dE_{\mathrm{sym}}/d\rho)_{\rho_0}$ of the symmetry energy around the saturation density $\rho_0$~\cite{Horowitz:2000xj,Furnstahl:2001un,Todd-Rutel:2005yzo,Centelles:2008vu,Zhang:2013wna,Xu:2020fdc}. In the past decades, the $\Delta r_{\mathrm{np}}$ has been measured experimentally through proton~\cite{Zenihiro:2010zz,Terashima:2008rb} and pion~\cite{Friedman:2012pa} scatterings, charge exchange reactions~\cite{Krasznahorkay:1999zz}, coherent pion photoproductions~\cite{Tarbert:2013jze}, and antiproton annihilations~\cite{Klos:2007is,Brown:2007zzc,Trzcinska:2001sy}. While the parity-violating electron-nucleus scatterings serve as a model-independent measurement, the data of the $\Delta r_{\mathrm{np}}$ for $^{208}$Pb~\cite{PREX:2021umo} and $^{48}$Ca~\cite{CREX:2022kgg} provided by the recent PREXII and CREX experiments favor respectively a large and small slope parameter $L$ of the $E_{sym}(\rho)$. More efforts are thus still needed to extract the $\Delta r_{\mathrm{np}}$ in neutron-rich nuclei, in order to put a more stringent constraint on the $E_{sym}(\rho)$.

It was realized that heavy-ion collisions could be an alternative way of measuring the $\Delta r_{\mathrm{np}}$. In relativistic heavy-ion collisions, the $\Delta r_{\mathrm{np}}$ in colliding nuclei can be extracted from mid-rapidity observables such as particle multiplicities or collective flows~\cite{Li:2019kkh,Giacalone:2023cet}, or from forward- and backward-rapidity observables such as the free neutron-to-proton yield ratio~\cite{Liu:2022kvz,Liu:2022xlm,Liu:2023qeq,Liu:2023rap}. In low- and intermediate-energy heavy-ion collisions, various probes of the $\Delta r_{\mathrm{np}}$ in colliding nuclei at midrapidites and forward rapidities have been proposed mostly by the quantum molecular dynamics approach~\cite{Sun:2009wf,Dai:2014rja,Dai:2015dua,Ding:2024jfr,Yang:2023bwm}. On the experimental side, while there are extensive measurements of isovector probes at midrapidities, observables at forward rapidities are available as well, e.g., at LNS-INFN by FAZIA Collaboration~\cite{FAZIA:2024vfo}, SARA and TAMU~\cite{Hagel:1994zz}, MSU~\cite{Mocko:2006tt}, and GSI~\cite{Pawlowski:2023gen} (see also Ref.~\cite{Ma:2021nwr}). The neutron removal reactions with a high collision energy has been proposed as a useful way of measuring the $\Delta r_{\mathrm{np}}$~\cite{PhysRevLett.119.262501}. In peripheral heavy-ion collisions, the $\Delta r_{\mathrm{np}}$ in colliding nuclei may also affect the $\pi^-/\pi^+$ yield ratio~\cite{Wei:2013sfa,Hartnack:2018sih}.

Generally, isovector oberavables in heavy-ion collisions, including those mentioned above, are affected by not only the $\Delta r_{\mathrm{np}}$ in colliding nuclei, but also the symmetry potential $U_{sym}$ in the heavy-ion collision dynamics, which characterizes the different potentials felt by neutrons and protons in neutron-rich matter and is related to the $E_{sym}(\rho)$. As a representative example, the free neutron-to-proton yield ratio has been proposed as a probe of the $\Delta r_{\mathrm{np}}$~\cite{Sun:2009wf,Yang:2023bwm}, but it is also a probe of the $U_{sym}$ in the dynamics of heavy-ion collisions~\cite{Li:1997rc,Zhang:2007hmv,Li:2005by,Famiano:2006rb}, and its kinetic energy spectrum is actually related to the neutron-proton effective mass splitting~\cite{Rizzo:2005mk,Kong:2015rla,Coupland:2014gya,Morfouace:2019jky}. While it is consistent to use a uniform $E_{sym}(\rho)$ in both the initialization of colliding nuclei and transport simulations of collision dynamics~\cite{Yang:2023bwm}, it is useful to distinguish the effects of the $\Delta r_{\mathrm{np}}$ in colliding nuclei and the symmetry potential in the collision dynamics on final observables, including those at midrapidities and forward rapidities, and search for robust probes of the $\Delta r_{\mathrm{np}}$ which are insensitive to $U_{sym}$. The present paper serves as such an investigation by using the IBUU transport model.
\vspace{-0.5cm}
\section{theoretical framework}
\label{theory}

In this section, we will briefly review the improved momentum-dependent nuclear effective interaction (ImMDI)~\cite{Xu:2014cwa} used in the present study, and the IBUU transport framework as well as the way to obtain the information of free nucleons in the simulation of intermediate-energy heavy-ion collisions.

\subsection{Effective nuclear interaction}

We start from an effective in-medium interaction between two nucleons at $\vec{r}_1$ and $\vec{r}_2$ with a zero-range density-dependent term and a Yukawa-type finite-range term, i.e.,
\begin{eqnarray}
v(\vec{r}_1,\vec{r}_2) &=& \frac{1}{6}t_3(1+x_3 P_\sigma)
\rho^\gamma\left(\frac{\vec{r}_1+\vec{r}_2}{2}\right)
\delta(\vec{r}_1-\vec{r}_2) \notag\\
&+& (W+G P_\sigma - H P_\tau - M P_\sigma P_\tau) \frac{e^{-\mu
|\vec{r}_1-\vec{r}_2|}}{|\vec{r}_1-\vec{r}_2|},\notag\\
\label{MDIyuk}
\end{eqnarray}
where $P_\sigma$ and $P_\tau$ are the spin- and isospin-exchange operators, respectively. In the Hartree-Fock approximation, the above interaction leads to the potential energy density in the form of~\cite{Das:2002fr,Xu:2010ce}
\begin{eqnarray}
V(\rho ,\delta ) &=&\frac{A_{u}\rho _{n}\rho _{p}}{\rho _{0}}+\frac{A_{l}}{%
2\rho _{0}}(\rho _{n}^{2}+\rho _{p}^{2})+\frac{B}{\sigma +1}\frac{\rho
^{\sigma +1}}{\rho _{0}^{\sigma }}  \notag \\
&\times &(1-x\delta ^{2})+\frac{1}{\rho _{0}}\sum_{\tau ,\tau ^{\prime
}}C_{\tau ,\tau ^{\prime }}  \notag \\
&\times &\int \int d^{3}pd^{3}p^{\prime }\frac{f_{\tau }(\vec{r},\vec{p}%
)f_{\tau ^{\prime }}(\vec{r},\vec{p}^{\prime
})}{1+(\vec{p}-\vec{p}^{\prime })^{2}/\Lambda ^{2}}. \label{MDIV}
\end{eqnarray}%
In the mean-field approximation, Eq.~(\ref{MDIV}) leads to the following
single-particle potential for nucleons with momentum $\vec{p}$
\begin{eqnarray}
U_\tau(\rho ,\delta ,\vec{p}) &=&A_{u}\frac{\rho _{-\tau }}{\rho _{0}}%
+A_{l}\frac{\rho _{\tau }}{\rho _{0}}  \notag \\
&+&B\left(\frac{\rho }{\rho _{0}}\right)^{\sigma }(1-x\delta ^{2})-4\tau x\frac{B}{%
\sigma +1}\frac{\rho ^{\sigma -1}}{\rho _{0}^{\sigma }}\delta \rho _{-\tau }
\notag \\
&+&\frac{2C_l}{\rho _{0}}\int d^{3}p^{\prime }\frac{f_{\tau }(%
\vec{r},\vec{p}^{\prime })}{1+(\vec{p}-\vec{p}^{\prime })^{2}/\Lambda ^{2}}
\notag \\
&+&\frac{2C_u}{\rho _{0}}\int d^{3}p^{\prime }\frac{f_{-\tau }(%
\vec{r},\vec{p}^{\prime })}{1+(\vec{p}-\vec{p}^{\prime })^{2}/\Lambda ^{2}}.
\label{MDIU}
\end{eqnarray}%
In the above, $\rho_{n(p)}$ is the neutron (proton) number density, $\rho=\rho_n+\rho_p$ and $\delta=(\rho_n-\rho_p)/\rho$ are the total nucleon density and the isospin asymmetry, respectively, and $\rho_0=0.16$ fm$^{-3}$ is the saturation density. $f_\tau(\vec{r},\vec{p})$ is the nucleon phase-space distribution function, with $\tau=1(-1)$ for neutrons (protons) being the isospin index. The eight parameters ($A_l$, $A_u$, $x$, $B$, $\sigma$, $C_l=C_{\tau,\tau}$,
$C_u=C_{\tau,-\tau}$, $\Lambda$) in the energy-density functional [Eq.~(\ref{MDIV})] are related to those ($t_3$, $x_3$, $\gamma$, $W$, $G$,
$H$, $M$, and $\mu$) in the effective nuclear interaction [Eq.~(\ref{MDIyuk})], and they are determined by empirical nuclear matter properties~\cite{Xu:2014cwa}. In ImMDI, the parameters $A_l$,
$A_u$, $C_l$, and $C_u$ are further expressed as
\begin{eqnarray}
A_{l}(x,y)&=&A_{l0} + y + x\frac{2B}{\sigma +1},  \label{AlImMDI}\\
A_{u}(x,y)&=&A_{u0} - y - x\frac{2B}{\sigma +1}, \label{AuImMDI}\\
C_{l}(y,z)&=&C_{l0} - 2y\frac{p^2_{f0}}{\Lambda^2\ln [(4 p^2_{f0} + \Lambda^2)/\Lambda^2]}, \label{ClImMDI}\\
C_{u}(y,z)&=&C_{u0} + 2y\frac{p^2_{f0}}{\Lambda^2\ln [(4 p^2_{f0} + \Lambda^2)/\Lambda^2]}, \label{CuImMDI}
\end{eqnarray}
where $p_{f0}$ is the nucleon Fermi momentum in symmetric nuclear matter at saturation density, and parameters $x$ and $y$ can be varied to adjust the values of the slope parameter $L$ of the symmetry energy and the momentum dependence of the symmetry potential $U_{sym}(p)$, with other macroscopic quantities fixed at their empirical values. Since the in-medium nucleon effective p-mass is defined as $\frac{m_{n(p)}^{*}}{m}=\left( 1+\frac{m}{p}\frac{\partial U_{n\left ( p \right )}}{\partial p}\right) ^{-1}$, with $m$ being the nucleon bare mass, $U_{sym}(p)=[U_n(p)-U_p(p)]/\delta$ is related to the isospin splitting of the nucleon effective mass in neutron-rich matter (see Ref.~\cite{Li:2018lpy} for a review).

\begin{figure}[ht]
  \centering
  \includegraphics[scale=0.3]{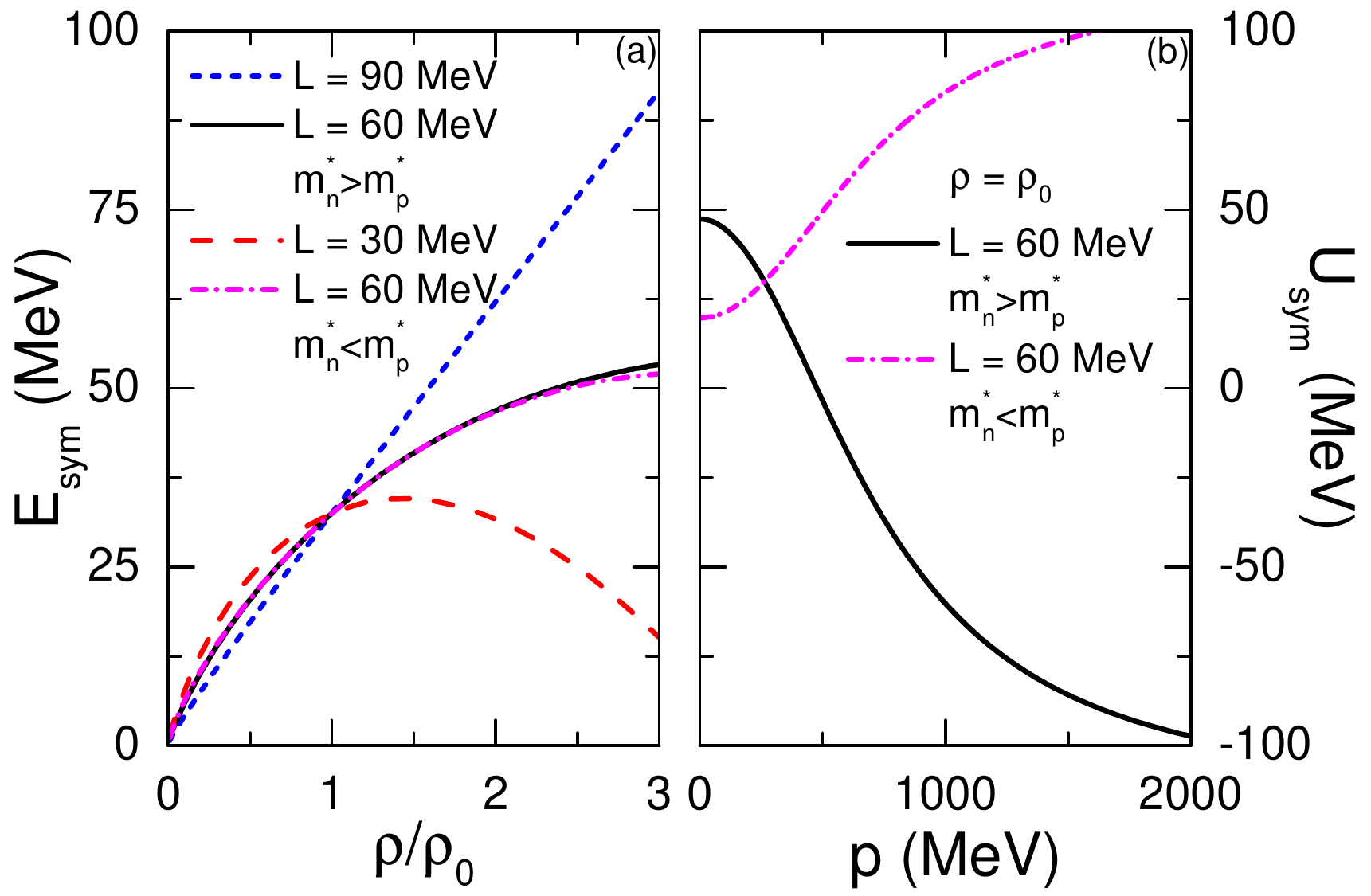}
  \caption{Density dependence of the nuclear symmetry energy $E_{sym}(\rho)$ (a) and the momentum dependence of the symmetry potential $U_{sym}(p)$ at $\rho_0$ (b) with different values of $x$ and $y$ in ImMDI.}
  \label{EsymUsym}
\end{figure}

In the present study, we compare the effect of $L$ as well as the momentum dependence of $U_{sym}(p)$ with that of the initial $\Delta r_{\mathrm{np}}$ in colliding nuclei on the final observables. The used $E_{sym}(\rho)$ and $U_{sym}(p)$ representing the uncertainty of the mean-field potential in intermediate-energy heavy-ion collisions are displayed in Fig.~\ref{EsymUsym}. By setting ($x=0.585$, $y=-115$ MeV), ($x=-0.016$, $y=-115$ MeV), and ($x=-0.618$, $y=-115$ MeV), we obtain different $E_{sym}(\rho)$ with $L=30$, 60, and 90 MeV, representing the uncertainty of $L=58.7 \pm 28.1$ MeV~\cite{Oertel:2016bki,Li:2013ola} from surveying dozens of analyses, while keeping the same value of $E_{sym}(\rho_0)$. For the effect of $U_{sym}(p)$, we compare the cases of ($x=-0.016$, $y=-115$ MeV) and ($x=0.999$, $y=115$ MeV), which lead to almost the same $E_{sym}(\rho)$ but considerably different $U_{sym}(p)$ and isospin splittings of the nucleon effective mass. We note here that we actually choose two extremely different momentum dependencies of $U_{sym}(p)$, corresponding to $(m_n^\star-m_p^\star)/m \approx 0.42\delta$ and $-0.25\delta$ at Fermi momentum in normal nuclear matter.

\subsection{Transport simulation}

For the simulation of intermediate-energy heavy-ion collisions, we basically solve the IBUU equation expressed as
\begin{eqnarray}
  \label{eq:boltz}
  \frac{\partial f_\tau(\vec{r},\vec{p})}{\partial t}
  +\left(\frac{\vec{p}}{\sqrt{m^2+\vec{p}^2}}+\frac{\partial U_\tau}{\partial \vec{p}}\right)\cdot\frac{\partial f_\tau(\vec{r},\vec{p})}{\partial \vec{r}} -\frac{\partial U_\tau}{\partial \vec{r}}\cdot\frac{\partial f_\tau(\vec{r},\vec{p})}{\partial \vec{p}}
  =I_c. \notag\\
\end{eqnarray}
In the above, $U_\tau$ is the mean-field potential, and $I_c$ represents the collision term with Pauli blocking.

For the mean-field evolution part, we use the lattice Hamiltonian method~\cite{Lenk:1989zz}. For a system with a total nucleon number $A$ and $N_{TP}$ test particles for each nucleon, the phase-space distribution function $f_{L,\tau}$ and the density $\rho_{L,\tau}$ at the sites of a three-dimensional cubic lattice for nucleons with isospin $\tau$ are
\begin{eqnarray}
f_{L,\tau}(\vec{r}_{\alpha},\vec{p}) &\sim& \sum_{i \in \tau}^{AN_{TP}}S(\vec{r}_{\alpha}-\vec{r}_i)\delta(\vec{p}-\vec{p}_{i}),\\
\rho_{L,\tau}(\vec{r}_{\alpha}) &=& \sum_{i \in \tau}^{AN_{TP}}S(\vec{r}_{\alpha}-\vec{r}_i).
\end{eqnarray}
In the above, $\alpha$ is the site index, $\vec{r}_{\alpha}$ is the position of the site $\alpha$, and $S$ is the shape function describing the contribution of a test particle at $\vec{r}_i$ to the value of the quantity at $\vec{r}_{\alpha}$, i.e.,
\begin{eqnarray}
S(\vec{r})=\frac{1}{N_{TP}(nl)^6}g(x)g(y)g(z)
\end{eqnarray}
with
\begin{eqnarray}
g(q)=(nl-|q|)\Theta(nl-|q|).
\end{eqnarray}
$l$ is the lattice spacing, $n$ determines the range of $S$, and $\Theta$ is the Heaviside function. We adopt the values of $l=1$ fm and $n=2$ in the present study. The Hamiltonian of the system can be expressed as
\begin{equation}\label{htotal}
H=\sum_{i=1}^{AN_{TP}}\sqrt{\vec{p}_{i}^{2}+m^2}+N_{TP}\widetilde{V},
\end{equation}
with $\widetilde{V}$ being the total potential energy expressed as
\begin{equation}
\widetilde{V}=l^3\sum_{\alpha}(V^{\rm ImMDI}_{\alpha} + V^{\rm grad}_{\alpha} + V^{\rm coul}_{\alpha}),
\end{equation}
where
\begin{eqnarray}
V^{\rm ImMDI}_{\alpha} &=& \frac{A_{u}\rho _{L,n}(\vec{r}_{\alpha})\rho _{L,p}(\vec{r}_{\alpha})}{\rho _{0}}+\frac{A_{l}}{2\rho _{0}}[\rho _{L,n}^{2}(\vec{r}_{\alpha})             \notag \\
& & +\rho _{L,p}^{2}(\vec{r}_{\alpha})]+\frac{B}{\sigma+1}\frac{\rho_{L}^{\sigma +1}(\vec{r}_{\alpha})}{\rho _{0}^{\sigma }}[1-x\delta_L ^{2}(\vec{r}_{\alpha})]+\frac{1}{\rho _{0}}                       \notag \\
& & \times \sum_{i,j} \sum_{\tau_{i} ,\tau_{j}}C_{\tau_{i} ,\tau_{j}}
	\frac{S(\vec{r}_{\alpha}-\vec{r}_i)S(\vec{r}_{\alpha}-\vec{r}_j)}{1+(\vec{p}_{i}
	-\vec{p}_{j})^{2}/\Lambda ^{2}}, \label{vimmdi}
\end{eqnarray}
\begin{equation}
V^{\rm grad}_{\alpha} = \frac{G_S}{2} [\nabla \rho_L(\vec{r}_{\alpha})]^2 - \frac{G_V}{2} \{\nabla [\rho_{L,n}(\vec{r}_{\alpha})-\rho_{L,p}(\vec{r}_{\alpha})]\}^2,\label{vgrad}
\end{equation}
and
\begin{eqnarray}
V^{\rm coul}_{\alpha} = \frac{e^2}{2}l^3 \sum_{\alpha^\prime} \frac{\rho_{L,p}(\vec{r}_{\alpha})\rho_{L,p}(\vec{r}_{\alpha^\prime})}{|\vec{r}_\alpha - \vec{r}_{\alpha^\prime}|} - \frac{3}{4} e^2 \left[\frac{3\rho_{L,p}(\vec{r}_\alpha)}{\pi} \right]^{4/3}\notag\\\label{vcoul}
\end{eqnarray}
are the corresponding contributions of the ImMDI energy-density functional, the density gradient interaction, and the Coulomb interaction, respectively. $\delta_L(\vec{r}_{\alpha})=[\rho _{L,n}(\vec{r}_{\alpha})-\rho _{L,p}(\vec{r}_{\alpha})]/[\rho _{L,n}(\vec{r}_{\alpha})+\rho _{L,p}(\vec{r}_{\alpha})]$ is the isospin asymmetry at $\vec{r}_{\alpha}$ with $\rho _{L,n}(\vec{r}_{\alpha})$ and $\rho _{L,p}(\vec{r}_{\alpha})$ being respectively the number density of neutrons and protons there. For the density gradient interaction, we use $G_S=132$ MeV fm$^5$ and $G_V=5$ MeV fm$^5$ from Ref.~\cite{Chen:2010qx}. The equations of motion for the $i$th test particle from the above Hamiltonian can thus be written as
\begin{eqnarray}
 \frac{d\vec{r}_{i}}{dt}&=&\frac{\partial H}{\partial\vec{p}_{i}}
 = \frac{\vec{p}_i}{\sqrt{\vec{p}_{i}^{2}+m^2}} + N_{TP}\frac{\partial\widetilde{V}}{\partial\vec{p}_{i}},  \label{rt}\\
 \frac{d\vec{p}_{i}}{dt} &=&-\frac{\partial H}{\partial\vec{r}_{i}}
 = -N_{TP}\frac{\partial\widetilde{V}}{\partial\vec{r}_{i}}.  \label{pt}
\end{eqnarray}

For the collision part, we consider nucleon-nucleon (NN) elastic collisions with the in-medium NN cross section $\sigma_{med}$ calculated from~\cite{Li:2005jy,Pandharipande:1992zz}
\begin{equation}
\sigma_{med} = \left(\frac{\mu_{NN}^\star}{\mu_{NN}}\right)^2\sigma_{vac},
\end{equation}
where $\mu_{NN}^{(\star)}=m_1^{(\star)} m_2^{(\star)}/(m_1^{(\star)}+m_2^{(\star)})$ is the reduced bare (effective) mass for the nucleon pair in the collision, and the neutron-neutron (proton-proton) and neutron-proton cross sections $\sigma_{vac}$ in vacuum are taken from the parameterized forms based on the phase-shift analysis~\cite{PhysRevC.15.1002,PhysRevC.96.044618}. For the energy and polar angular dependence of these isospin-dependent NN cross sections in free space, see, e.g., Fig. 1 in Ref.~\cite{PhysRevC.109.034614}. For inelastic channels, we have incorporated the isospin-dependent $N+N \leftrightarrow N+\Delta$ and $\Delta \leftrightarrow N+\pi$ processes, with the cross sections and decay widths taken from Ref.~\cite{TMEP:2023ifw}, and the isospin-dependent $N+N \leftrightarrow N+N^\star$ and $N^\star \leftrightarrow N+\pi$ processes, with the cross sections and decay widths taken from Ref.~\cite{Li:1995pra}. The isospin-dependent Pauli blocking probabilities for nucleons in final states of collisions and decays are evaluated from the occupation number in local phase-space cells with the interpolation method.

In order to explore the effect of $\Delta r_{\mathrm{np}}$ in colliding nuclei on final observables, we sample the coordinates of initial neutrons and protons according to the density distribution generated from spherical Skyrme-Hartree-Fock (SHF) calculations, for which we used the MSL0 force~\cite{Chen:2010qx}. With the standard SHF energy-density functional, we are able to express the values of ten parameters in terms of ten macroscopic quantities~\cite{Chen:2010qx}, and this allows us to adjust only the value of the slope parameter $L$ of the $E_{sym}(\rho)$ while fixing the values of other physics quantities to get different $\Delta r_{\mathrm{np}}$. In the present study, we set $L=30$, 60, and 90 MeV in the SHF calculation, and they lead to $\Delta r_{\mathrm{np}}=0.16$, 0.19, and 0.23 fm in $^{124}$Sn. The momenta of neutrons (protons) are generated within the Fermi sphere according to the local neutron (proton) density, and they are then boosted according to the beam energy to perform the simulation of a heavy-ion collision.

We basically use the free neutron-to-proton yield ratio as final observables in intermediate-energy heavy-ion collisions. To obtain the information of final free nucleons, we do IBUU simulations until about $t=80$ fm/c when the dynamics is mostly finished. Afterwards, heavy clusters with $A>3$ are grouped by nucleons close in phase space with $\Delta r < 3$ fm and $\Delta p <300$ MeV/c~\cite{Li:1997rc}, and the formation of deuterons, tritons, and $^3$He is described by a Wigner function approach~\cite{Chen:2003ava,Sun:2017ooe}. The rest nucleons are the directly produced free nucleons. On the other hand, a considerable amount of nucleons can be produced from the deexcitation of heavy clusters, and this is especially so for nucleons produced at forward rapidities from the deexcitation of the spectator matter in peripheral collisions. We use the GEMINI model~\cite{Charity:1988zz,Charity:2010wk} to describe the deexcitation process, for which the excitation energy is calculated by subtracting the ground-state energy~\cite{Wang:2021xhn} from the total energy of the fragment, with the latter calculated based on the test-particle method using a simplified SHF-like energy density functional~\cite{Liu:2022xlm}, i.e.,
\begin{eqnarray}\label{E}
E &=& \frac{1}{N_{TP}}\sum_{i} \left(\sqrt{m^2+p_i^2}-m\right) \notag\\
&+& \int d^3r \left[\frac{a}{2} \left(\frac{\rho}{\rho_0}\right)^2  +  \frac{b}{\sigma+1}\left(\frac{\rho}{\rho_0}\right)^{\sigma+1}\right] \notag\\
&+& \int d^3r E_{\mathrm{sym}}^{\mathrm{pot}} \left(\frac{\rho}{\rho_0}\right)^{\gamma} \frac{(\rho_n-\rho_p)^2}{\rho} \notag\\
&+& \int d^3r \left\{ \frac{G_S}{2} (\nabla \rho)^2 - \frac{G_V}{2} [\nabla (\rho_n-\rho_p)]^2\right\} \notag\\
&+& \frac{e^2}{2} \int d^3r d^3r' \frac{\rho_p(\vec{r})\rho_p(\vec{r}')}{|\vec{r}-\vec{r}'|} - \frac{3e^2}{4} \int d^3r \left( \frac{3\rho_p}{\pi}\right)^{4/3}, \notag\\
\end{eqnarray}
with $G_S$ and $G_V$ set as the same in the Eq.~(\ref{vgrad}), and parameters $a$, $b$, $\sigma$, $E_{\mathrm{sym}}^{\mathrm{pot}}$, and $\gamma$ adjusted to mimic the same nuclear matter properties as in ImMDI.

\section{results and discussions}
\label{results}

In this section, we investigate the sensitivities of the free neutron-to-proton yield ratios at both midrapidities and forward rapidities to the initial $\Delta r_{\mathrm{np}}$ in colliding nuclei and the nucleon mean-field potential in both peripheral and central collisions. We will first explore $^{124}$Sn+$^{124}$Sn collisions at $E_{beam}=600A$ MeV as in the GSI experiment~\cite{Pawlowski:2023gen}, and then move to a higher energy of $E_{beam}=2A$ GeV, since we expect an increasing nucleus structure effect and a decreasing mean-field potential effect in the dynamics of collisions at higher energies.

\subsection{Peripheral collisions}

\begin{figure}[ht]
  \centering
  \includegraphics[scale=0.3]{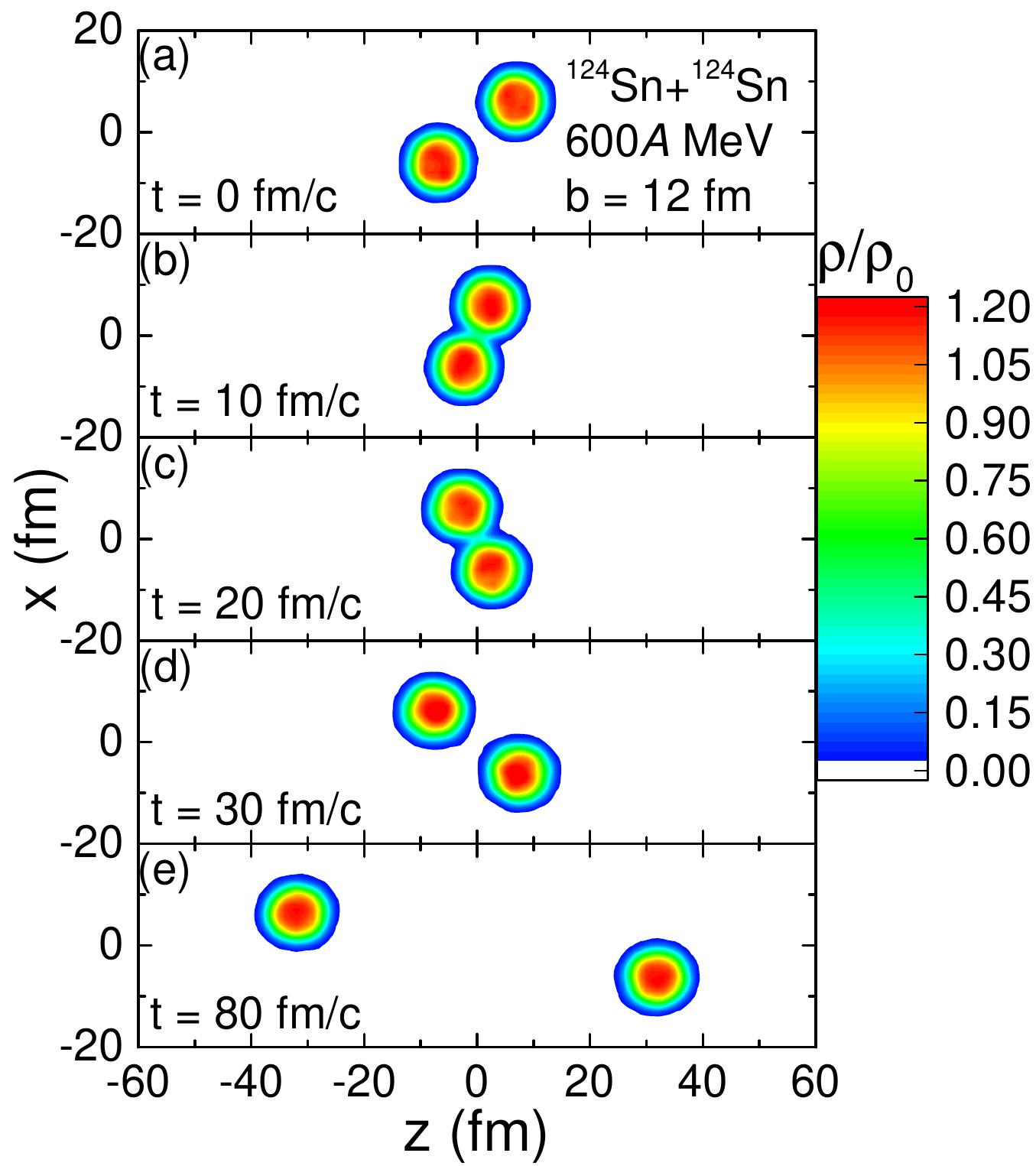}
  \caption{Density contours in the reaction plane (x-o-z plane) for peripheral $^{124}$Sn+$^{124}$Sn collisions at $E_{beam}=600A$ MeV at $t=0$ (a), 10 (b), 20 (c), 30 (d), and 80 (e) fm/c from IBUU simulation with 200 test particles.}
  \label{den_E600b12}
\end{figure}

\begin{figure}[ht]
  \centering
  \includegraphics[scale=0.32]{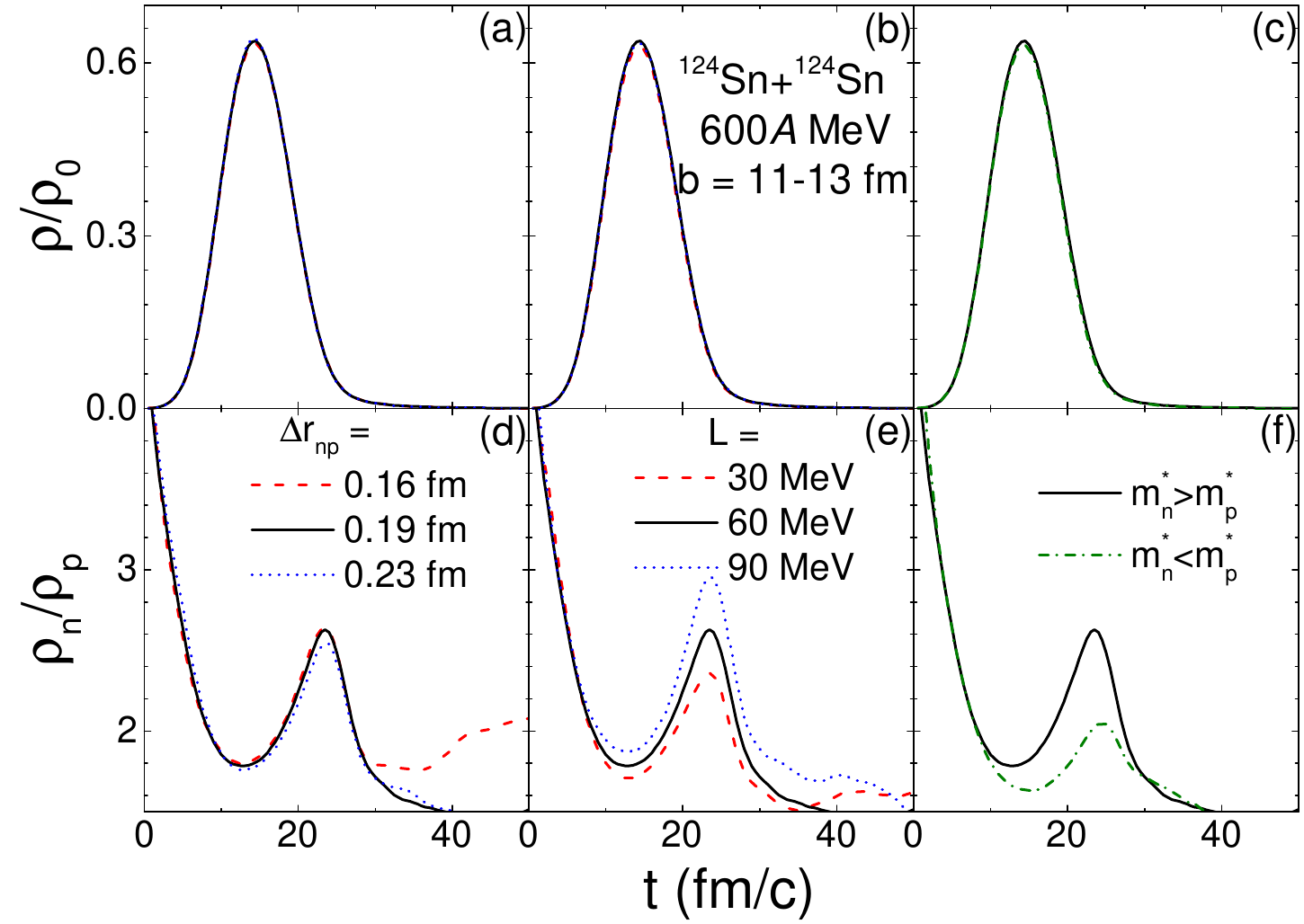}
  \caption{Time evolution of the central density (upper) and neutron-proton asymmetry (lower) in peripheral $^{124}$Sn+$^{124}$Sn collisions at $E_{beam}=600A$ MeV from different initial neutron-skin thickness in $^{124}$Sn (left column), different symmetry energies in the collision dynamics (middle column), and different neutron-proton effective mass splittings in the collision dynamics (right column). Here the neutron-proton asymmetry is calculated by taking the ratio of event-averaged central neutron density to proton density.}
  \label{rhodelta_E600b12}
\end{figure}

We begin by investigating peripheral $^{124}$Sn+$^{124}$Sn collisions at $600A$ MeV. The density evolution in the reaction plane from a default case of IBUU simulations, i.e., with initial $\Delta r_{np}=0.19$ fm and parameter values of ($x=-0.016$, $y=-115$
MeV), is displayed in Fig.~\ref{den_E600b12}, and the time evolutions of the central density and the neutron-proton asymmetry for the same centrality range in the default parameterization and by varying the initial $\Delta r_{np}$ or the $U_{sym}$ are compared in Fig.~\ref{rhodelta_E600b12}. At the initial stage, we put the two nuclei close enough so that the sampled density distribution with a desired $\Delta r_{np}$ can be maintained before collision. Later on, the two nuclei rub their edges and skim each other. In peripheral collisions, the participant matter is formed by the surface in colliding nuclei, and the central density reaches about $0.6\rho_0$, which is almost the same for different cases. It is seen that the initial $\Delta r_{np}$, which is obtained from SHF calculations with reasonable nuclear forces, does not affect the central neutron-proton asymmetry by much, compared to the effect of the $U_{sym}$. For a stiffer (softer) $E_{sym}$ in the collision dynamics, the participant matter with a low density becomes more (less) neutron-rich. On the other hand, the case of $m_n^\star>m_p^\star$ leads to a more neutron-rich participant matter compared to the case of $m_n^\star<m_p^\star$, since neutrons has a higher effective mass than protons in the former case and they tend to stay in the system rather than to be emitted out. Here we already see a larger effect from the $U_{sym}$ compared to that from the initial $\Delta r_{np}$. While free nucleons are emitted during the collision stage, the two spectator matters pass away, and after some relaxation time they become stable. We stop the simulation at $t=80$ fm/c, and have tested that the results of interest are insensitive to the cut-off time.

\begin{figure}[ht]
  \centering
  \includegraphics[scale=0.3]{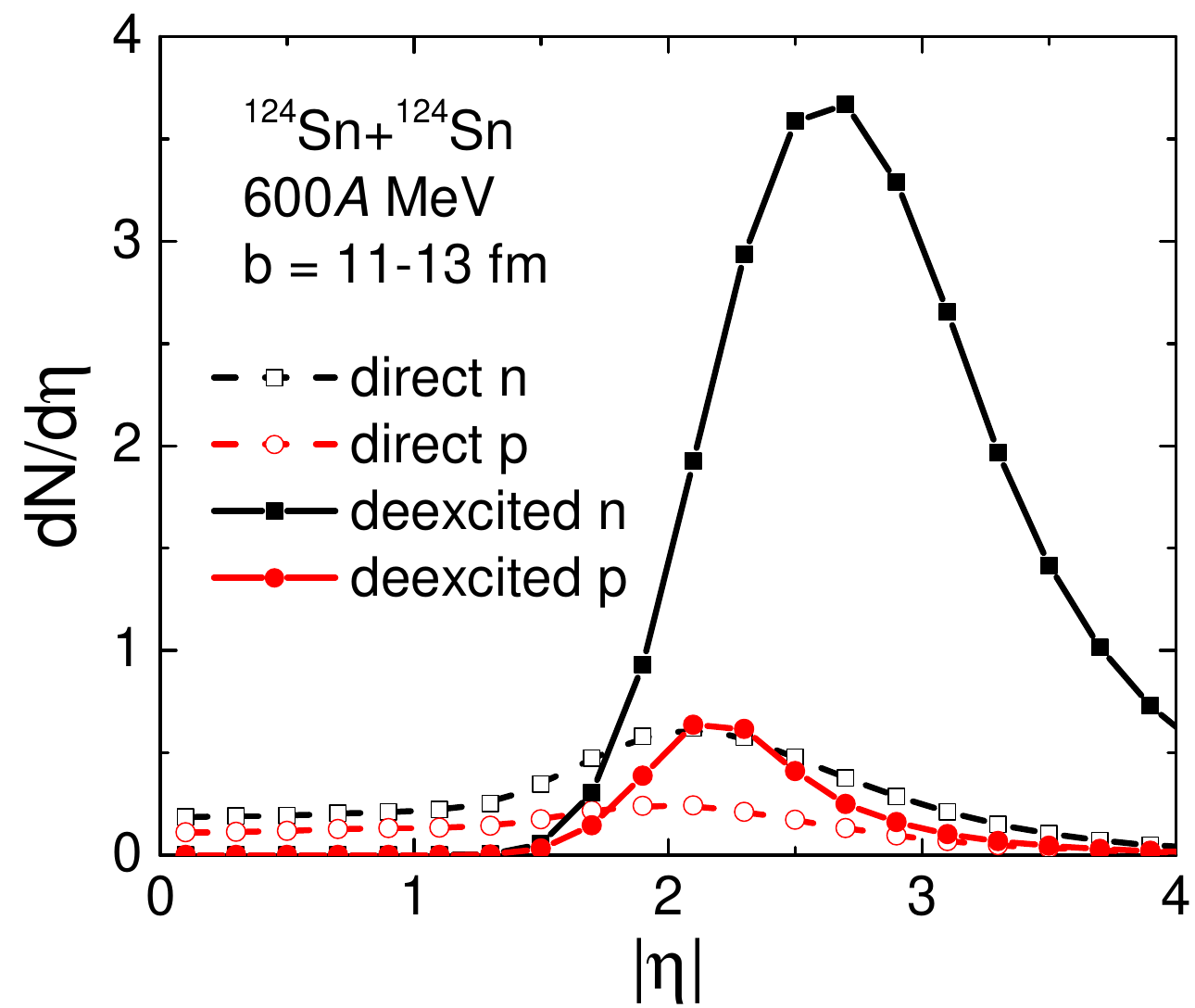}
  \caption{Pseudorapidity dependence of neutrons and protons from direct production [direct n(p)] and from deexcitation [deexcited n(p)] in peripheral $^{124}$Sn+$^{124}$Sn collisions at $E_{beam}=600A$ MeV.}
  \label{np_eta_E600b12}
\end{figure}

The pseudorapidity dependencies of free neutrons and protons from different production mechanisms for the default case as Fig.~\ref{den_E600b12}, which are symmetric in forward and backward rapidities in the symmetric collision system here, are compared in Fig.~\ref{np_eta_E600b12}. One mechanism is the direct production, i.e., from the emission during the collision dynamics, and these nucleons are produced in the whole pseudorapidity range including midpseudorapidities. Another mechanism is from the deexcitation of residue fragments, as seen from the bottom panel in Fig.~\ref{den_E600b12}, and these nucleons are produced only at large pseudorapidities. In peripheral collisions, as investigated here, about $80\%$ free nucleons are from the deexcitation of the residue fragments and only about $20\%$ free nucleons are from direct production. We also found that about $85\%$ free neutrons and $60\%$ free protons are from the deexcitation of residue fragments, compared to the total numbers of free neutrons and protons, respectively. Here we see that the neutron-to-proton yield ratio from deexcitation is much larger than that from direct production.

\begin{figure}[ht]
  \centering
  \includegraphics[scale=0.4]{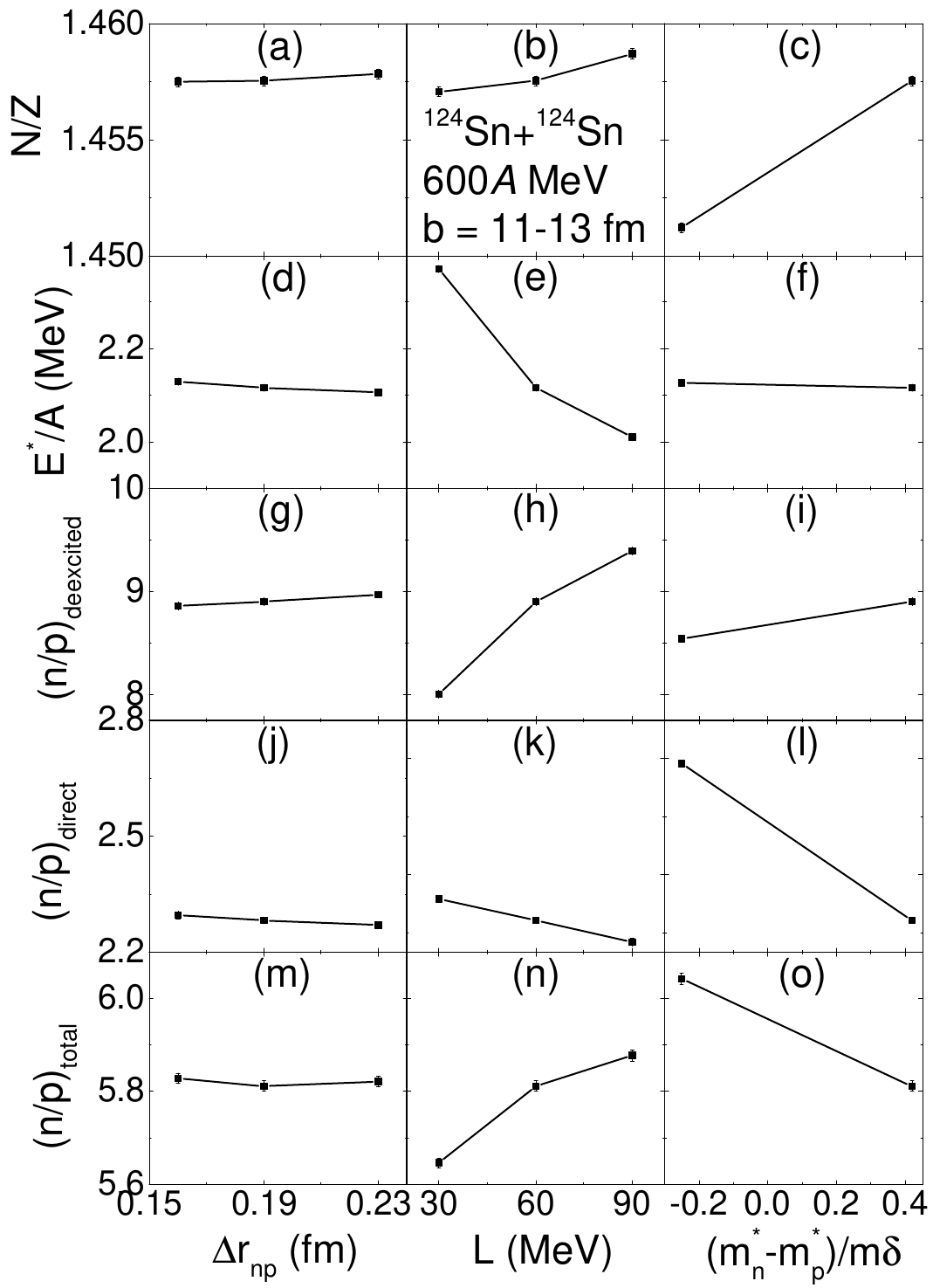}
  \caption{Dependence of the $N/Z$ (first row) and excitation energy per nucleon $E^\star/A$ (second row) in residue nuclei as well as the free neutron-to-proton yield ratio from deexcitation (third row), direction production (fourth row), and both (bottom row) in peripheral $^{124}$Sn+$^{124}$Sn collisions at $E_{beam}=600A$ MeV on the initial neutron-skin thickness in $^{124}$Sn (left column), the slope parameter of the symmetry energy in the collision dynamics (middle column), and the neutron-proton effective mass splitting in the collision dynamics (right column). }
  \label{np_E600b12}
\end{figure}

The ratios of neutron-to-proton number $N/Z$ in residue fragments after collisions from different initial $\Delta r_{np}$ and $U_{sym}$ are compared in the first row of Fig.~\ref{np_E600b12}. The $N/Z$ ratio is rather insensitive to the initial $\Delta r_{np}$, but increases with the increasing stiffness of the $E_{sym}$, and is larger for $m_n^\star>m_p^\star$ than that for $m_n^\star<m_p^\star$. These are consistent with the behaviors of the central neutron-proton asymmetry shown in the second row of Fig.~\ref{rhodelta_E600b12}. The excitation energy per nucleon $E^\star/A$ in residue fragments, as shown in the second row of Fig.~\ref{np_E600b12}, is seen to be rather insensitive to the initial $\Delta r_{np}$ and the neutron-proton effective mass splitting, but decreases with the increasing stiffness of the $E_{sym}$. The latter is due to the symmetry energy term in Eq.~(\ref{E}), where we adjust the value of the $\gamma$ parameter to reproduce the same slope parameter $L$ of the $E_{sym}(\rho)$ in the collision dynamics. Since the average density of residue fragments is below $\rho_0$, a stiffer $E_{sym}$ leads to a lower $E^\star/A$ and thus a smaller number of free nucleons from the deexcitation process. This generally leads to a larger free neutron-to-proton yield ratio from the deexcitation process $(n/p)_{deexcited}$, as shown in the third row of Fig.~\ref{np_E600b12}. The $(n/p)_{deexcited}$ yield ratio is also larger with a larger $N/Z$ ratio in residue fragments, as can be seen from results with different neutron-proton effective mass splittings. For the free neutron-to-proton yield ratio from the direct production $(n/p)_{direct}$, as shown in the fourth row of Fig.~\ref{np_E600b12}, it is smaller for a stiffer $E_{sym}$ corresponding to a weaker $U_{sym}$ at subsaturation densities, and is smaller for $m_n^\star>m_p^\star$ than for $m_n^\star<m_p^\star$ due to the suppression of neutron emission in the former case. It is seen that the $E_{sym}$ has an opposite effect on the $(n/p)_{deexcited}$ and the $(n/p)_{direct}$. The total $n/p$ yield ratio is seen to be larger for a stiffer $E_{sym}$ dominated by $(n/p)_{deexcited}$, and smaller for $m_n^\star>m_p^\star$ than for $m_n^\star<m_p^\star$ dominated by $(n/p)_{direct}$. We note that the total neutron number at $|\eta|>3.13$, corresponding to the forward polar angle range of $5^o$, is about 7.5, similar to the value from the GSI data in Ref.~\cite{Pawlowski:2023gen}, showing that the deexcitation process is treated reasonably in our study. Compared to the effect of the $U_{sym}$ on the total $n/p$ yield ratio, the effect of the initial $\Delta r_{np}$ is much smaller in peripheral collisions.

\subsection{Central collisions}

\begin{figure}[ht]
  \centering
  \includegraphics[scale=0.3]{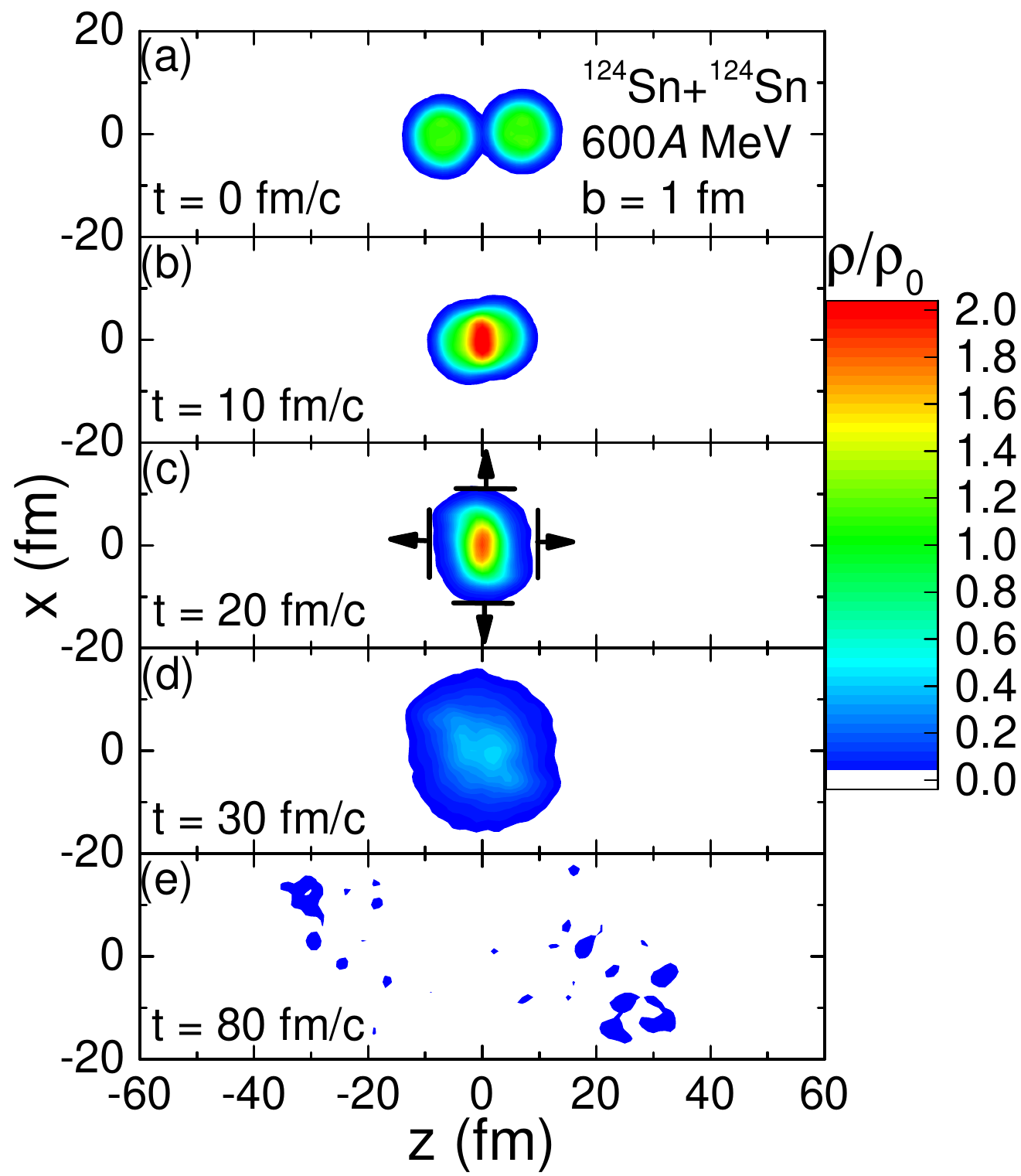}
  \caption{Density contours in the reaction plane (x-o-z plane) for central $^{124}$Sn+$^{124}$Sn collisions at $E_{beam}=600A$ MeV at $t=0$ (a), 10 (b), 20 (c), 30 (d), and 80 (e) fm/c from IBUU simulation with 200 test particles. At $t=20$ fm/c, both the expansions in the transverse plane (x-o-y plane) and in the longitudinal direction are marked.}
  \label{den_E600b1}
\end{figure}

\begin{figure}[ht]
  \centering
  \includegraphics[scale=0.32]{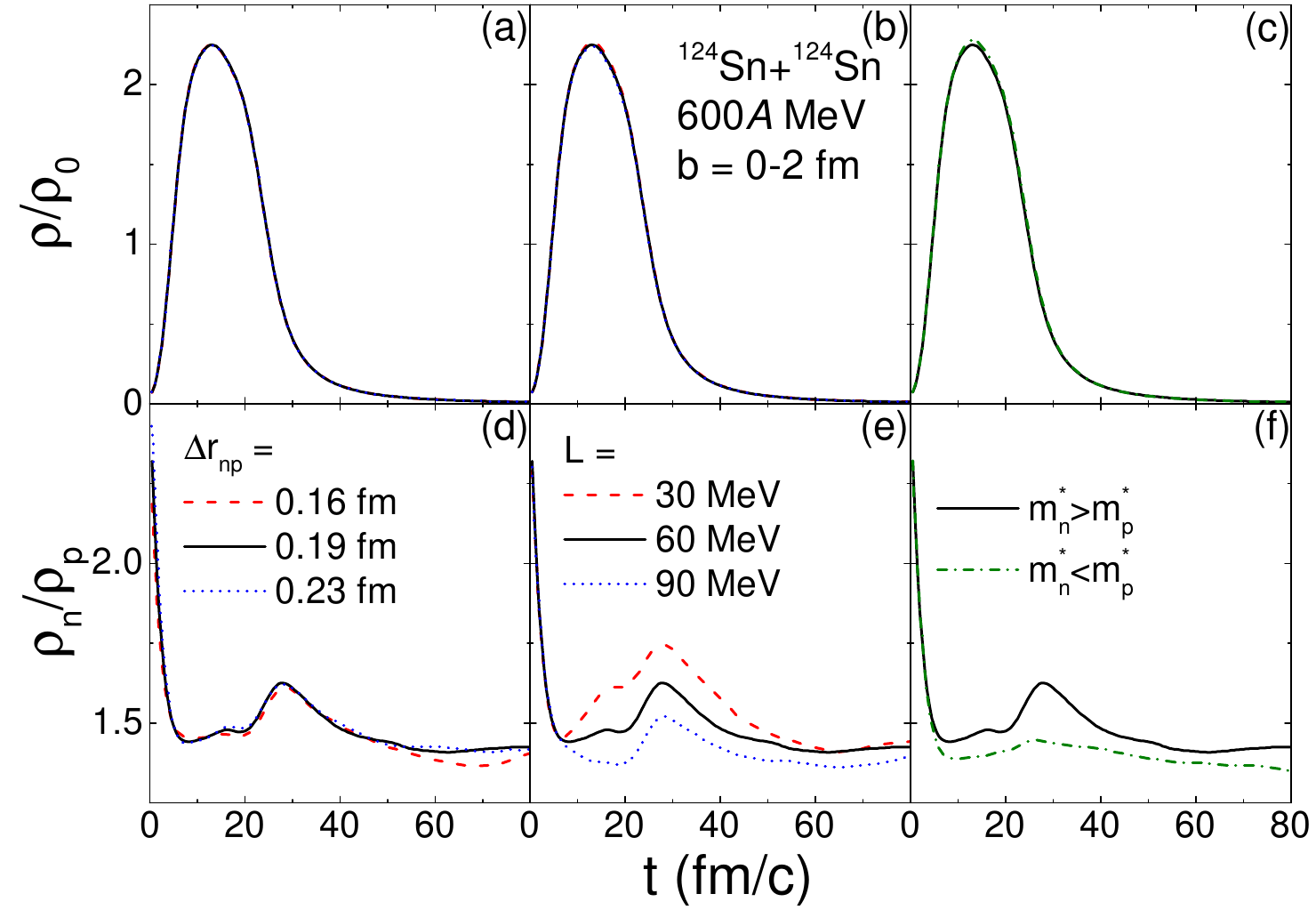}
  \caption{Time evolution of the central density (upper) and neutron-proton asymmetry (lower) in central $^{124}$Sn+$^{124}$Sn collisions at $E_{beam}=600A$ MeV from different initial neutron-skin thicknesses in $^{124}$Sn (left column), different symmetry energies in the collision dynamics (middle column), and different neutron-proton effective mass splittings in the collision dynamics (right column). }
  \label{rhodelta_E600b1}
\end{figure}

As shown in the previous subsection, the free neutron-to-proton yield ratio is more sensitive to the $U_{sym}$ but less sensitive to the initial $\Delta r_{np}$ in colliding nuclei in peripheral collisions. Referencing the previous studies~\cite{Liu:2022kvz,Liu:2022xlm,Liu:2023qeq,Liu:2023rap} by some of us, we now investigate the situation in central $^{124}$Sn+$^{124}$Sn collisions at $600A$ MeV. The density evolution in the reaction plane from the default case of IBUU simulations is displayed in Fig.~\ref{den_E600b1}, and the time evolutions of the central density and neutron-proton asymmetry for different initial $\Delta r_{np}$ and $U_{sym}$ are displayed in Fig.~\ref{rhodelta_E600b1}. It is seen that the participant matter reaches a much higher density in the collision stage compared to the case of peripheral collisions, and the maximum density is again insensitive to the initial $\Delta r_{np}$ or $U_{sym}$. The central neutron-proton asymmetry is larger for a larger $L$ corresponding to the larger $E_{sym}$ at suprasaturation densities, and larger for $m_n^\star>m_p^\star$ than for $m_n^\star<m_p^\star$ due to a weaker neutron emission in the former case, but is still insensitive to the initial $\Delta r_{np}$ during the compression stage. Later on, the system expands both in the transverse plane and in the longitudinal direction, as marked in the plot, and the surface of such expansion contains low-density neutron-rich matter generated by the initial $\Delta r_{np}$ in colliding nuclei and the mean-field dynamics. At $t=80$ fm/c, no bulk matter is observed, except that there are some nucleons in the low-density phase moving in the forward and backward rapidities, and we hope that they may carry some information of the initial $\Delta r_{np}$ in colliding nuclei as what we found in relativistic heavy-ion collisions~\cite{Liu:2022kvz,Liu:2022xlm,Liu:2023qeq,Liu:2023rap}.

\begin{figure}[ht]
  \centering
  \includegraphics[scale=0.5]{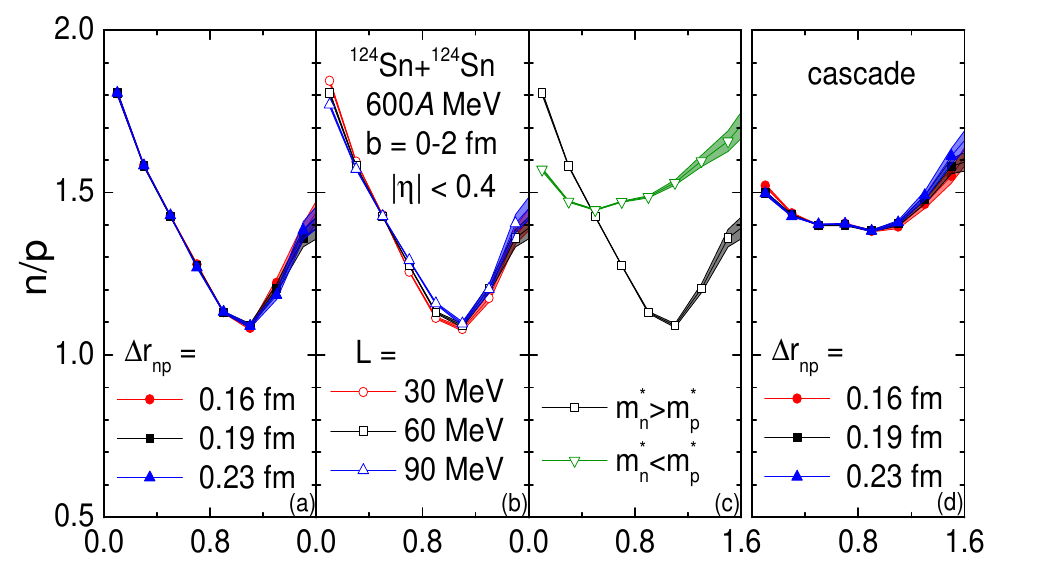}\\
  \includegraphics[scale=0.5]{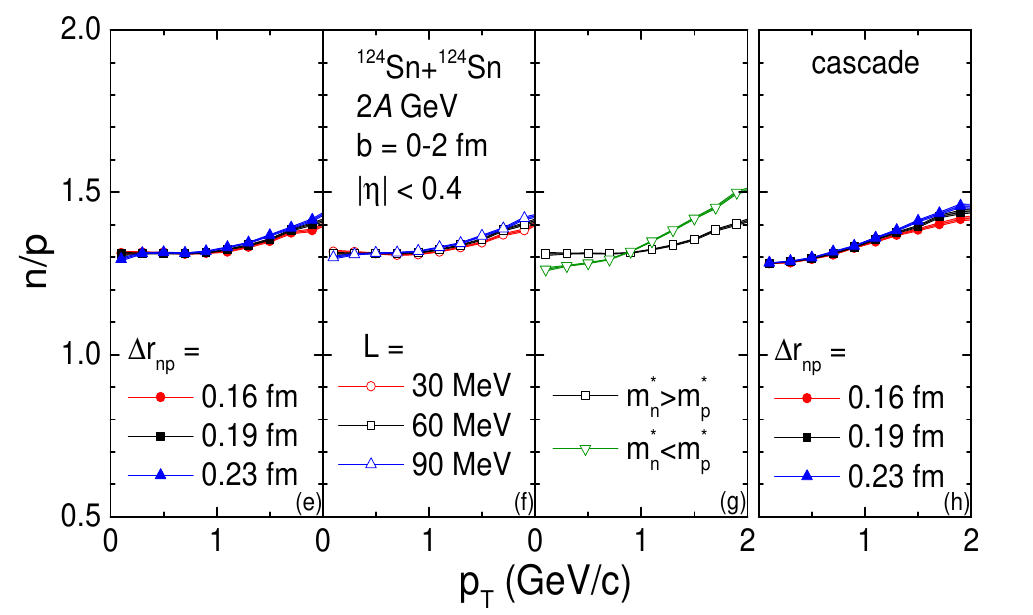}
  \caption{Transverse momentum dependence of the free neutron-to-proton yield ratio at midpseudorapidities $|\eta|<0.4$ in central $^{124}$Sn+$^{124}$Sn collisions at $600A$ MeV (upper panel) and $2A$ GeV (lower panel) from different initial neutron-skin thicknesses in $^{124}$Sn (first column), different symmetry energies in the collision dynamics (second column), and different neutron-proton effective mass splittings in the collision dynamics (third column). Results from cascade calculations without mean-field potential or Coulomb interaction using different initial neutron-skin thicknesses are shown in the fourth column.}
  \label{pt_b1}
\end{figure}

The transverse momentum dependence of the free neutron-to-proton yield ratio at midpseudorapidities in central $^{124}$Sn+$^{124}$Sn collisions from different initial $\Delta r_{np}$ and $U_{sym}(p)$ are compared in Fig.~\ref{pt_b1}. At $600A$ MeV as shown in the upper panel of Fig.~\ref{pt_b1}, $\Delta r_{np}$ is seen to have almost no effect on the results. A stiffer (softer) $E_{sym}$ leads to a larger (smaller) neutron-to-proton yield ratio $n/p$ at larger $p_T$, since neutrons (protons) are affected by a more (less) repulsive force in the high-density phase. For the case of $m_n^\star<m_p^\star$, the energetic neutrons (protons) feel a more (less) repulsive force compared to the case of $m_n^\star>m_p^\star$, as can be seen from Fig.~\ref{EsymUsym} (b), and this significantly enhances the $n/p$ yield ratio at high $p_T$. The effects of $E_{sym}$ and the isospin splitting of the nucleon effective mass on the free $n/p$ yield ratio are consistent with those found in previous studies~\cite{Li:1997rc,Zhang:2007hmv,Li:2005by,Famiano:2006rb,Rizzo:2005mk,Kong:2015rla,Coupland:2014gya,Morfouace:2019jky}. We have also analyzed the results at $2A$ GeV, as shown in the lower panel of Fig.~\ref{pt_b1}. Here we observe a larger $n/p$ yield ratio at high $p_T$ from a larger initial $\Delta r_{np}$, likely due to the emission of the surface neutron-rich matter in the transverse plane pushed by the expansion of the central participant matter (see the transverse expansion marked in Fig.~\ref{den_E600b1} (c) for illustration). On the other hand, the effects of the $E_{sym}$ and the isospin splitting of nucleon effective mass remain qualitatively similar at this collision energy while quantitatively smaller compared with those at $600A$ MeV, and they are, respectively, similar to and larger than the effect from $\Delta r_{np}$, making it impossible to extract $\Delta r_{np}$ free from the uncertainty of $U_{sym}(p)$. We have also compared in the right column results from cascade calculations without mean-field potential or Coulomb interaction using different initial $\Delta r_{np}$, and in this case all free nucleons are used in the analysis. While the effect is still small, it shows the ``maximum'' effect on the $n/p$ yield ratio from the initial $\Delta r_{np}$ regardless of $U_{sym}(p)$.

\begin{figure}[ht]
  \centering
  \includegraphics[scale=0.5]{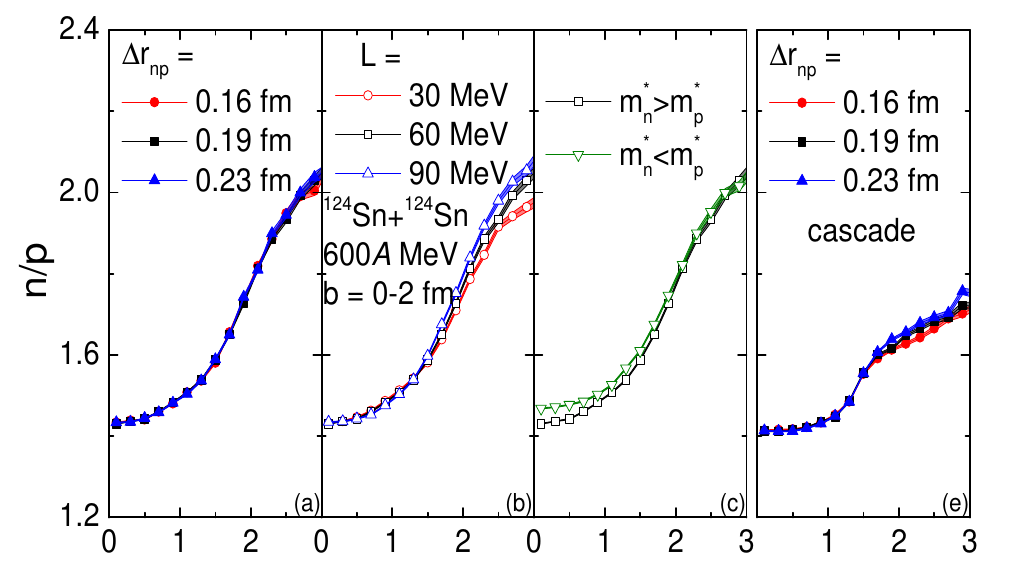}
  \includegraphics[scale=0.5]{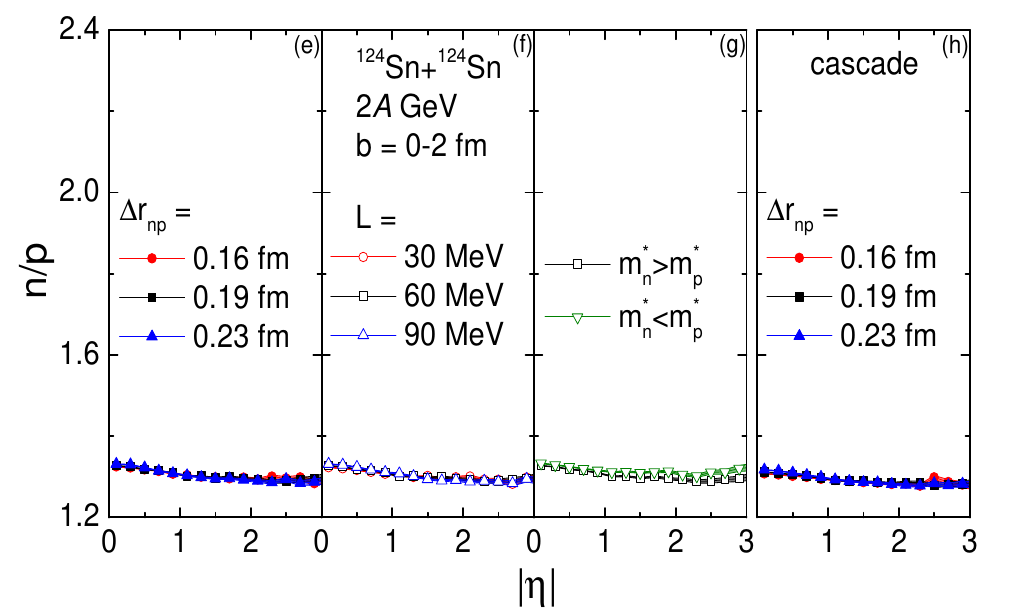}
  \caption{Pseudorapidity dependence of the free neutron-to-proton yield ratio in central $^{124}$Sn+$^{124}$Sn collisions at $600A$ MeV (upper panel) and $2A$ GeV (lower panel) from different initial neutron-skin thicknesses in $^{124}$Sn (first column), different symmetry energies in the collision dynamics (second column), and different neutron-proton effective mass splittings in the collision dynamics (third column). Results from cascade calculations without mean-field potential or Coulomb interaction using different initial neutron-skin thicknesses are shown in the fourth column.}
  \label{eta_b1}
\end{figure}

We now move to the investigation at large rapidities, and for the symmetric $^{124}$Sn+$^{124}$Sn collision system we take the averaged results in forward and backward rapidities. Figure~\ref{eta_b1} displays the pseudorapidity dependence of the free $n/p$ yield ratio in central $^{124}$Sn+$^{124}$Sn collisions for different cases as in Fig.~\ref{pt_b1}. The $n/p$ yield ratio shows almost no sensitivity to the initial $\Delta r_{np}$ in colliding nuclei at both $600A$ MeV and $2A$ GeV, as shown in the left column of Fig.~\ref{eta_b1}. At large $|\eta|$, the $n/p$ yield ratio is larger for a stiffer $E_{sym}$ at $600A$ MeV, since a considerable amount of these nucleons come from the low-density region, which is more neutron-rich for a stiffer $E_{sym}$. This effect is, however, rather weak at $2A$ GeV. Since the average nucleon momenta are higher than about 300 MeV/c, the average $U_{sym}(p)$ is larger for $m_n^\star<m_p^\star$ than that for $m_n^\star>m_p^\star$ (see Fig.~\ref{EsymUsym} (b)), and this leads to a larger $n/p$ yield ratio in the whole $|\eta|$ range for the former case. Results from cascade calculations shown in the right column also indicate that the effect of $n/p$ yield ratio at large $|\eta|$ from initial $\Delta r_{np}$ is generally smaller than that from $U_{sym}(p)$.

\begin{figure}[ht]
  \centering
  \includegraphics[scale=0.5]{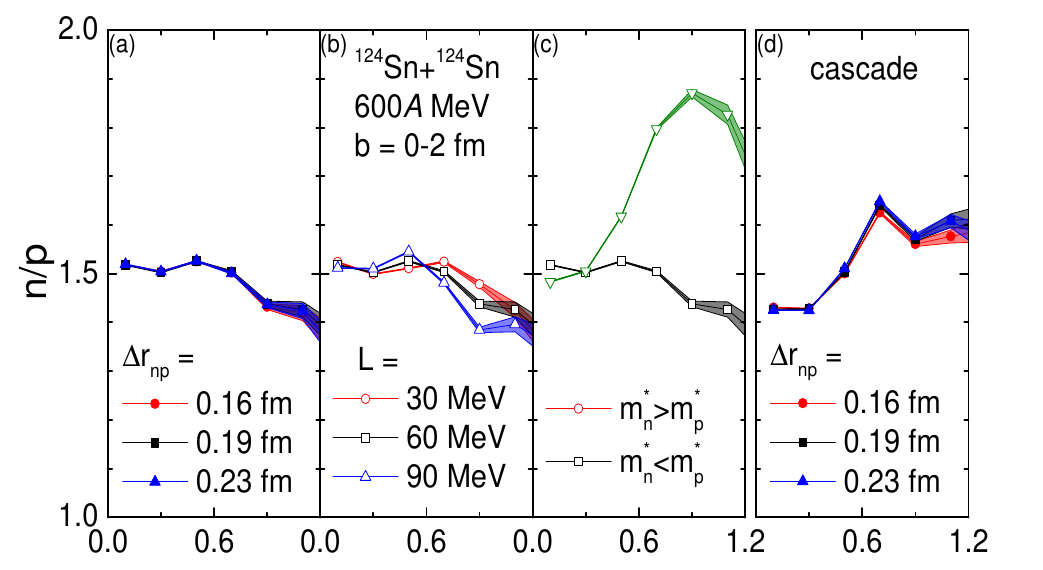}
  \includegraphics[scale=0.5]{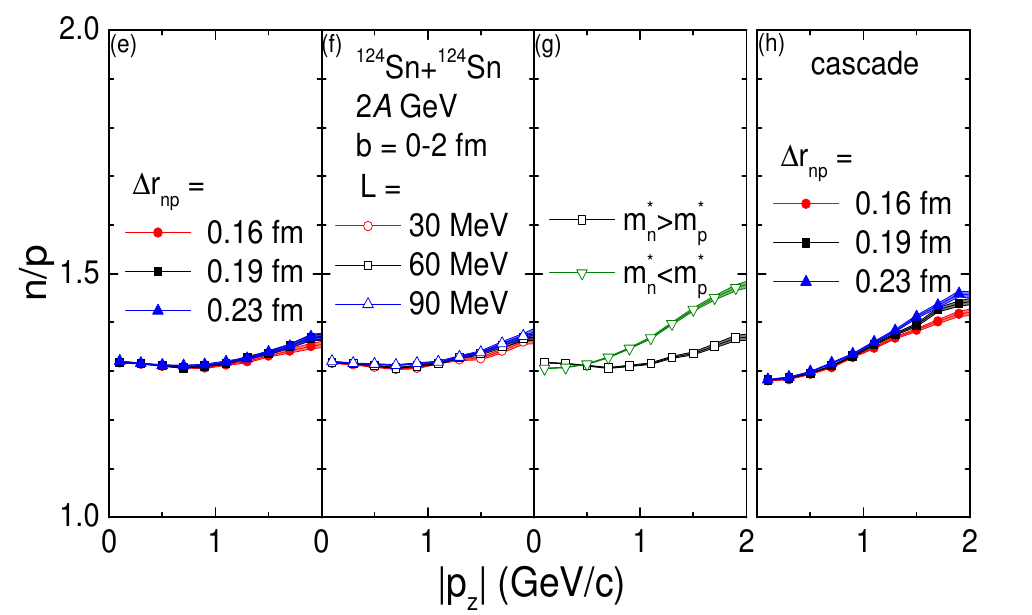}
  \caption{Longitudinal momentum dependence of the free neutron-to-proton yield ratio in central $^{124}$Sn+$^{124}$Sn collisions at $600A$ MeV (upper panel) and $2A$ GeV (lower panel) from different initial neutron-skin thicknesses in $^{124}$Sn (first column), different symmetry energies in the collision dynamics (second column), and different neutron-proton effective mass splittings in the collision dynamics (third column). Results from cascade calculations without mean-field potential or Coulomb interaction using different initial neutron-skin thicknesses are shown in the fourth column.}
  \label{pz_b1}
\end{figure}

\begin{table*}[t]
\centering
\caption{$n/p$ yield ratios at $p_T>1$ GeV/c and $|p_z|>1$ GeV/c in central $^{124}$Sn+$^{124}$Sn collisions at $2A$ GeV from different initial neutron-skin thicknesses $\Delta r_{np}$ in $^{124}$Sn, different slope parameters $L$ of the symmetry energy in the collision dynamics, and different neutron-proton effective mass splittings $(m_n^\star-m_p^\star)/m\delta$ in the collision dynamics.}
\label{tab1}
%\renewcommand\arraystretch{1.5}
%\setlength{\tabcolsep}{5mm}
% \resizebox{80mm}{!}
\begin{tabular}{|c|c|c|c|}
\hline
$\Delta r_{np}$ (fm) & 0.16 & 0.19 & 0.23\\
\hline
$n/p (p_T>1~\text{GeV/c})$ & $ 1.3445 \pm 0.0010 $ & $ 1.3521 \pm 0.0010 $ & $ 1.3613 \pm 0.0010 $ \\
$n/p (|p_z|>1~\text{GeV/c})$ & $ 1.3194 \pm 0.0013 $ & $ 1.3246 \pm 0.0013 $ & $ 1.3266 \pm 0.0013 $ \\
\hline\hline
$L$ (MeV) & 30 & 60 & 90\\
\hline
$n/p (p_T>1~\text{GeV/c})$ & $ 1.3434 \pm 0.0010 $ & $ 1.3521 \pm 0.0010 $ & $ 1.3601 \pm 0.0010 $ \\
$n/p (|p_z|>1~\text{GeV/c})$ & $  1.3212 \pm 0.0013 $ & $  1.3246 \pm 0.0013 $ & $ 1.3285 \pm 0.0013 $ \\
\hline\hline
$(m_n^\star-m_p^\star)/m\delta$ & $0.42$ & & $-0.25$  \\
\hline
$n/p (p_T>1~\text{GeV/c})$ & $ 1.3521 \pm 0.0010 $ & & $ 1.4026 \pm 0.0011 $  \\
$n/p (|p_z|>1~\text{GeV/c})$ &  $  1.3246 \pm 0.0013 $ & & $   1.3935 \pm 0.0014 $   \\
\hline
\end{tabular}
\end{table*}

Instead of analyzing the pseudorapidity dependence, we compare in Fig.~\ref{pz_b1} the longitudinal momentum dependence of the free $n/p$ yield ratio in the same cases as in Fig.~\ref{eta_b1}. While again the results show no sensitivity to the initial $\Delta r_{np}$ at $600A$ MeV, as shown in Fig.~\ref{pz_b1} (a), the $n/p$ yield ratio at large $|p_z|$ is seen to be larger for a larger initial $\Delta r_{np}$ at $2A$ GeV. At large $|p_z|$ instead of large $|\eta|$, we think that the $n/p$ yield ratio is dominated by nucleons emitted from the high-density phase at $600A$ MeV but by nucleons emitted from the longitudinal surface of the participant matter (see the longitudinal expansion marked in Fig.~\ref{den_E600b1} (c) for illustration) at $2A$ GeV. This explains the insensitivity of $n/p$ yield ratio to the $\Delta r_{np}$ but a larger $n/p$ yield ratio at large $|p_z|$ for a softer $E_{sym}$ at $600A$ MeV, and the larger $n/p$ yield ratios at large $|p_z|$ for both a larger $\Delta r_{np}$ and a stiffer $E_{sym}$ at $2A$ GeV. The mean-field potential with $m_n^\star>m_p^\star$ ($m_n^\star<m_p^\star$) leads to larger (smaller) $n/p$ yield ratio at larger $|p_z|$ but smaller (larger) $n/p$ yield ratio at smaller $|p_z|$, consistent with the momentum dependence of $U_{sym}(p)$ shown in Fig.~\ref{EsymUsym} (b). Again, results from cascade calculations shown in the right column of Fig.~\ref{eta_b1} display the ``maximum'' effect of the initial $\Delta r_{np}$ on the $n/p$ yield ratio at large $|p_z|$.

Based on the discussions from Figs.~\ref{pt_b1}, \ref{eta_b1}, \ref{pz_b1}, we found that the largest effect on the $n/p$ yield ratio from the initial $\Delta r_{np}$ can be observed at large transverse momenta or large longitudinal momenta. The summarized results are listed in Table~\ref{tab1}. At $p_T>1$ GeV/c and $|p_z|>1$ GeV/c in central $^{124}$Sn+$^{124}$Sn collisions at $2A$ GeV, it is seen that the effect on the $n/p$ yield ratio from the initial $\Delta r_{np}$ is comparable to the $E_{sym}$ effect in the collision dynamics, although it is still much smaller than the effect of the neutron-proton effective mass splitting.

\section{conclusions}
\label{summary}

We have investigated the possibility of using the free neutron-to-proton yield ratio $n/p$ to extract the neutron-skin thickness $\Delta r_{np}$ in nuclei by their collisions at intermediate energies, and have analyzed extensively the $n/p$ yield ratio at both midrapidities and forward rapidities in peripheral and central $^{124}$Sn+$^{124}$Sn collisions based on the IBUU transport model. In peripheral collisions, the $n/p$ yield ratio is more sensitive to the symmetry energy which affects the excitation energy and the deexcitation of heavy fragments at forward rapidities, and the neutron-proton effective mass splitting which affects the isospin-dependent direct production of free nucleons. In central collisions, the $n/p$ yield ratios at midrapidities and large transverse momenta, those at forward rapidities, and those at large longitudinal momenta are analyzed. The resulting $n/p$ yield ratio is more sensitive to the symmetry potential in the collision dynamics than to the initial $\Delta r_{np}$ in colliding nuclei in most cases. While the $n/p$ yield ratio at large transverse or longitudinal momenta in central collisions shows the largest isospin effect, a reasonable range of $\Delta r_{np}$ leads to only about $1\%$ effect on the $n/p$ yield ratio at the collision energy of a few GeV/nucleon, smaller than the uncertainty due to the symmetry potential in the collision dynamics.

\begin{acknowledgments}
This work is supported by the Strategic Priority Research Program of the Chinese Academy of Sciences under Grant No. XDB34030000, the National Natural Science Foundation of China under Grant Nos. 12375125, 12035011, and 11975167, and the Fundamental Research Funds for the Central Universities.
\end{acknowledgments}

\bibliography{IBUU_nskin}

\begin{thebibliography}{68}%
\makeatletter
\providecommand \@ifxundefined [1]{%
 \@ifx{#1\undefined}
}%
\providecommand \@ifnum [1]{%
 \ifnum #1\expandafter \@firstoftwo
 \else \expandafter \@secondoftwo
 \fi
}%
\providecommand \@ifx [1]{%
 \ifx #1\expandafter \@firstoftwo
 \else \expandafter \@secondoftwo
 \fi
}%
\providecommand \natexlab [1]{#1}%
\providecommand \enquote  [1]{``#1''}%
\providecommand \bibnamefont  [1]{#1}%
\providecommand \bibfnamefont [1]{#1}%
\providecommand \citenamefont [1]{#1}%
\providecommand \href@noop [0]{\@secondoftwo}%
\providecommand \href [0]{\begingroup \@sanitize@url \@href}%
\providecommand \@href[1]{\@@startlink{#1}\@@href}%
\providecommand \@@href[1]{\endgroup#1\@@endlink}%
\providecommand \@sanitize@url [0]{\catcode `\\12\catcode `\$12\catcode
  `\&12\catcode `\#12\catcode `\^12\catcode `\_12\catcode `\%12\relax}%
\providecommand \@@startlink[1]{}%
\providecommand \@@endlink[0]{}%
\providecommand \url  [0]{\begingroup\@sanitize@url \@url }%
\providecommand \@url [1]{\endgroup\@href {#1}{\urlprefix }}%
\providecommand \urlprefix  [0]{URL }%
\providecommand \Eprint [0]{\href }%
\providecommand \doibase [0]{http://dx.doi.org/}%
\providecommand \selectlanguage [0]{\@gobble}%
\providecommand \bibinfo  [0]{\@secondoftwo}%
\providecommand \bibfield  [0]{\@secondoftwo}%
\providecommand \translation [1]{[#1]}%
\providecommand \BibitemOpen [0]{}%
\providecommand \bibitemStop [0]{}%
\providecommand \bibitemNoStop [0]{.\EOS\space}%
\providecommand \EOS [0]{\spacefactor3000\relax}%
\providecommand \BibitemShut  [1]{\csname bibitem#1\endcsname}%
\let\auto@bib@innerbib\@empty
%</preamble>
\bibitem [{\citenamefont {Li}\ \emph {et~al.}(2014)\citenamefont {Li},
  \citenamefont {Ramos}, \citenamefont {Verde},\ and\ \citenamefont
  {Vidana}}]{Li:2014oda}%
  \BibitemOpen
  \bibfield  {author} {\bibinfo {author} {\bibfnamefont {Bao-An}\ \bibnamefont
  {Li}}, \bibinfo {author} {\bibfnamefont {Angels}\ \bibnamefont {Ramos}},
  \bibinfo {author} {\bibfnamefont {Giuseppe}\ \bibnamefont {Verde}}, \ and\
  \bibinfo {author} {\bibfnamefont {Isaac}\ \bibnamefont {Vidana}},\ }\bibfield
   {title} {\enquote {\bibinfo {title} {{Topical issue on nuclear symmetry
  energy}},}\ }\href {\doibase 10.1140/epja/i2014-14009-x} {\bibfield
  {journal} {\bibinfo  {journal} {Eur. Phys. J. A}\ }\textbf {\bibinfo {volume}
  {50}},\ \bibinfo {pages} {9} (\bibinfo {year} {2014})}\BibitemShut {NoStop}%
\bibitem [{\citenamefont {Steiner}\ \emph {et~al.}(2005)\citenamefont
  {Steiner}, \citenamefont {Prakash}, \citenamefont {Lattimer},\ and\
  \citenamefont {Ellis}}]{Steiner:2004fi}%
  \BibitemOpen
  \bibfield  {author} {\bibinfo {author} {\bibfnamefont {Andrew~W.}\
  \bibnamefont {Steiner}}, \bibinfo {author} {\bibfnamefont {Madappa}\
  \bibnamefont {Prakash}}, \bibinfo {author} {\bibfnamefont {James~M.}\
  \bibnamefont {Lattimer}}, \ and\ \bibinfo {author} {\bibfnamefont {Paul~J.}\
  \bibnamefont {Ellis}},\ }\bibfield  {title} {\enquote {\bibinfo {title}
  {{Isospin asymmetry in nuclei and neutron stars}},}\ }\href {\doibase
  10.1016/j.physrep.2005.02.004} {\bibfield  {journal} {\bibinfo  {journal}
  {Phys. Rept.}\ }\textbf {\bibinfo {volume} {411}},\ \bibinfo {pages}
  {325--375} (\bibinfo {year} {2005})},\ \Eprint
  {http://arxiv.org/abs/nucl-th/0410066} {arXiv:nucl-th/0410066} \BibitemShut
  {NoStop}%
\bibitem [{\citenamefont {Baran}\ \emph {et~al.}(2005)\citenamefont {Baran},
  \citenamefont {Colonna}, \citenamefont {Greco},\ and\ \citenamefont
  {Di~Toro}}]{Baran:2004ih}%
  \BibitemOpen
  \bibfield  {author} {\bibinfo {author} {\bibfnamefont {V.}~\bibnamefont
  {Baran}}, \bibinfo {author} {\bibfnamefont {M.}~\bibnamefont {Colonna}},
  \bibinfo {author} {\bibfnamefont {V.}~\bibnamefont {Greco}}, \ and\ \bibinfo
  {author} {\bibfnamefont {M.}~\bibnamefont {Di~Toro}},\ }\bibfield  {title}
  {\enquote {\bibinfo {title} {{Reaction dynamics with exotic beams}},}\ }\href
  {\doibase 10.1016/j.physrep.2004.12.004} {\bibfield  {journal} {\bibinfo
  {journal} {Phys. Rept.}\ }\textbf {\bibinfo {volume} {410}},\ \bibinfo
  {pages} {335--466} (\bibinfo {year} {2005})},\ \Eprint
  {http://arxiv.org/abs/nucl-th/0412060} {arXiv:nucl-th/0412060} \BibitemShut
  {NoStop}%
\bibitem [{\citenamefont {Li}\ \emph {et~al.}(2008)\citenamefont {Li},
  \citenamefont {Chen},\ and\ \citenamefont {Ko}}]{Li:2008gp}%
  \BibitemOpen
  \bibfield  {author} {\bibinfo {author} {\bibfnamefont {Bao-An}\ \bibnamefont
  {Li}}, \bibinfo {author} {\bibfnamefont {Lie-Wen}\ \bibnamefont {Chen}}, \
  and\ \bibinfo {author} {\bibfnamefont {Che~Ming}\ \bibnamefont {Ko}},\
  }\bibfield  {title} {\enquote {\bibinfo {title} {{Recent Progress and New
  Challenges in Isospin Physics with Heavy-Ion Reactions}},}\ }\href {\doibase
  10.1016/j.physrep.2008.04.005} {\bibfield  {journal} {\bibinfo  {journal}
  {Phys. Rept.}\ }\textbf {\bibinfo {volume} {464}},\ \bibinfo {pages}
  {113--281} (\bibinfo {year} {2008})},\ \Eprint
  {http://arxiv.org/abs/0804.3580} {arXiv:0804.3580 [nucl-th]} \BibitemShut
  {NoStop}%
\bibitem [{\citenamefont {Baldo}\ and\ \citenamefont
  {Burgio}(2016)}]{Baldo:2016jhp}%
  \BibitemOpen
  \bibfield  {author} {\bibinfo {author} {\bibfnamefont {M.}~\bibnamefont
  {Baldo}}\ and\ \bibinfo {author} {\bibfnamefont {G.~F.}\ \bibnamefont
  {Burgio}},\ }\bibfield  {title} {\enquote {\bibinfo {title} {{The nuclear
  symmetry energy}},}\ }\href {\doibase 10.1016/j.ppnp.2016.06.006} {\bibfield
  {journal} {\bibinfo  {journal} {Prog. Part. Nucl. Phys.}\ }\textbf {\bibinfo
  {volume} {91}},\ \bibinfo {pages} {203--258} (\bibinfo {year} {2016})},\
  \Eprint {http://arxiv.org/abs/1606.08838} {arXiv:1606.08838 [nucl-th]}
  \BibitemShut {NoStop}%
\bibitem [{\citenamefont {Horowitz}\ and\ \citenamefont
  {Piekarewicz}(2001)}]{Horowitz:2000xj}%
  \BibitemOpen
  \bibfield  {author} {\bibinfo {author} {\bibfnamefont {C.~J.}\ \bibnamefont
  {Horowitz}}\ and\ \bibinfo {author} {\bibfnamefont {J.}~\bibnamefont
  {Piekarewicz}},\ }\bibfield  {title} {\enquote {\bibinfo {title} {{Neutron
  star structure and the neutron radius of Pb-208}},}\ }\href {\doibase
  10.1103/PhysRevLett.86.5647} {\bibfield  {journal} {\bibinfo  {journal}
  {Phys. Rev. Lett.}\ }\textbf {\bibinfo {volume} {86}},\ \bibinfo {pages}
  {5647} (\bibinfo {year} {2001})},\ \Eprint
  {http://arxiv.org/abs/astro-ph/0010227} {arXiv:astro-ph/0010227} \BibitemShut
  {NoStop}%
\bibitem [{\citenamefont {Furnstahl}(2002)}]{Furnstahl:2001un}%
  \BibitemOpen
  \bibfield  {author} {\bibinfo {author} {\bibfnamefont {R.~J.}\ \bibnamefont
  {Furnstahl}},\ }\bibfield  {title} {\enquote {\bibinfo {title} {{Neutron
  radii in mean field models}},}\ }\href {\doibase
  10.1016/S0375-9474(02)00867-9} {\bibfield  {journal} {\bibinfo  {journal}
  {Nucl. Phys. A}\ }\textbf {\bibinfo {volume} {706}},\ \bibinfo {pages}
  {85--110} (\bibinfo {year} {2002})},\ \Eprint
  {http://arxiv.org/abs/nucl-th/0112085} {arXiv:nucl-th/0112085} \BibitemShut
  {NoStop}%
\bibitem [{\citenamefont {Todd-Rutel}\ and\ \citenamefont
  {Piekarewicz}(2005)}]{Todd-Rutel:2005yzo}%
  \BibitemOpen
  \bibfield  {author} {\bibinfo {author} {\bibfnamefont {B.~G.}\ \bibnamefont
  {Todd-Rutel}}\ and\ \bibinfo {author} {\bibfnamefont {J.}~\bibnamefont
  {Piekarewicz}},\ }\bibfield  {title} {\enquote {\bibinfo {title}
  {{Neutron-Rich Nuclei and Neutron Stars: A New Accurately Calibrated
  Interaction for the Study of Neutron-Rich Matter}},}\ }\href {\doibase
  10.1103/PhysRevLett.95.122501} {\bibfield  {journal} {\bibinfo  {journal}
  {Phys. Rev. Lett.}\ }\textbf {\bibinfo {volume} {95}},\ \bibinfo {pages}
  {122501} (\bibinfo {year} {2005})},\ \Eprint
  {http://arxiv.org/abs/nucl-th/0504034} {arXiv:nucl-th/0504034} \BibitemShut
  {NoStop}%
\bibitem [{\citenamefont {Centelles}\ \emph {et~al.}(2009)\citenamefont
  {Centelles}, \citenamefont {Roca-Maza}, \citenamefont {Vinas},\ and\
  \citenamefont {Warda}}]{Centelles:2008vu}%
  \BibitemOpen
  \bibfield  {author} {\bibinfo {author} {\bibfnamefont {M.}~\bibnamefont
  {Centelles}}, \bibinfo {author} {\bibfnamefont {X.}~\bibnamefont
  {Roca-Maza}}, \bibinfo {author} {\bibfnamefont {X.}~\bibnamefont {Vinas}}, \
  and\ \bibinfo {author} {\bibfnamefont {M.}~\bibnamefont {Warda}},\ }\bibfield
   {title} {\enquote {\bibinfo {title} {{Nuclear symmetry energy probed by
  neutron skin thickness of nuclei}},}\ }\href {\doibase
  10.1103/PhysRevLett.102.122502} {\bibfield  {journal} {\bibinfo  {journal}
  {Phys. Rev. Lett.}\ }\textbf {\bibinfo {volume} {102}},\ \bibinfo {pages}
  {122502} (\bibinfo {year} {2009})},\ \Eprint {http://arxiv.org/abs/0806.2886}
  {arXiv:0806.2886 [nucl-th]} \BibitemShut {NoStop}%
\bibitem [{\citenamefont {Zhang}\ and\ \citenamefont
  {Chen}(2013)}]{Zhang:2013wna}%
  \BibitemOpen
  \bibfield  {author} {\bibinfo {author} {\bibfnamefont {Zhen}\ \bibnamefont
  {Zhang}}\ and\ \bibinfo {author} {\bibfnamefont {Lie-Wen}\ \bibnamefont
  {Chen}},\ }\bibfield  {title} {\enquote {\bibinfo {title} {{Constraining the
  symmetry energy at subsaturation densities using isotope binding energy
  difference and neutron skin thickness}},}\ }\href {\doibase
  10.1016/j.physletb.2013.08.002} {\bibfield  {journal} {\bibinfo  {journal}
  {Phys. Lett. B}\ }\textbf {\bibinfo {volume} {726}},\ \bibinfo {pages}
  {234--238} (\bibinfo {year} {2013})},\ \Eprint
  {http://arxiv.org/abs/1302.5327} {arXiv:1302.5327 [nucl-th]} \BibitemShut
  {NoStop}%
\bibitem [{\citenamefont {Xu}\ \emph {et~al.}(2020)\citenamefont {Xu},
  \citenamefont {Xie},\ and\ \citenamefont {Li}}]{Xu:2020fdc}%
  \BibitemOpen
  \bibfield  {author} {\bibinfo {author} {\bibfnamefont {Jun}\ \bibnamefont
  {Xu}}, \bibinfo {author} {\bibfnamefont {Wen-Jie}\ \bibnamefont {Xie}}, \
  and\ \bibinfo {author} {\bibfnamefont {Bao-An}\ \bibnamefont {Li}},\
  }\bibfield  {title} {\enquote {\bibinfo {title} {{Bayesian inference of
  nuclear symmetry energy from measured and imagined neutron skin thickness in
  $^{116,118,120,122,124,130,132}$Sn, $^{208}$Pb , and $^{48}$Ca}},}\ }\href
  {\doibase 10.1103/PhysRevC.102.044316} {\bibfield  {journal} {\bibinfo
  {journal} {Phys. Rev. C}\ }\textbf {\bibinfo {volume} {102}},\ \bibinfo
  {pages} {044316} (\bibinfo {year} {2020})},\ \Eprint
  {http://arxiv.org/abs/2007.07669} {arXiv:2007.07669 [nucl-th]} \BibitemShut
  {NoStop}%
\bibitem [{\citenamefont {Zenihiro}\ \emph {et~al.}(2010)\citenamefont
  {Zenihiro} \emph {et~al.}}]{Zenihiro:2010zz}%
  \BibitemOpen
  \bibfield  {author} {\bibinfo {author} {\bibfnamefont {J.}~\bibnamefont
  {Zenihiro}} \emph {et~al.},\ }\bibfield  {title} {\enquote {\bibinfo {title}
  {{Neutron density distributions of Pb-204, Pb-206, Pb-208 deduced via proton
  elastic scattering at Ep=295 MeV}},}\ }\href {\doibase
  10.1103/PhysRevC.82.044611} {\bibfield  {journal} {\bibinfo  {journal} {Phys.
  Rev. C}\ }\textbf {\bibinfo {volume} {82}},\ \bibinfo {pages} {044611}
  (\bibinfo {year} {2010})}\BibitemShut {NoStop}%
\bibitem [{\citenamefont {Terashima}\ \emph {et~al.}(2008)\citenamefont
  {Terashima} \emph {et~al.}}]{Terashima:2008rb}%
  \BibitemOpen
  \bibfield  {author} {\bibinfo {author} {\bibfnamefont {S.}~\bibnamefont
  {Terashima}} \emph {et~al.},\ }\bibfield  {title} {\enquote {\bibinfo {title}
  {{Proton elastic scattering from tin isotopes at 295-MeV and systematic
  change of neutron density distributions}},}\ }\href {\doibase
  10.1103/PhysRevC.77.024317} {\bibfield  {journal} {\bibinfo  {journal} {Phys.
  Rev. C}\ }\textbf {\bibinfo {volume} {77}},\ \bibinfo {pages} {024317}
  (\bibinfo {year} {2008})},\ \Eprint {http://arxiv.org/abs/0801.3082}
  {arXiv:0801.3082 [nucl-ex]} \BibitemShut {NoStop}%
\bibitem [{\citenamefont {Friedman}(2012)}]{Friedman:2012pa}%
  \BibitemOpen
  \bibfield  {author} {\bibinfo {author} {\bibfnamefont {E.}~\bibnamefont
  {Friedman}},\ }\bibfield  {title} {\enquote {\bibinfo {title} {{Neutron skins
  of $^{208}$Pb and $^{48}$Ca from pionic probes}},}\ }\href {\doibase
  10.1016/j.nuclphysa.2012.09.007} {\bibfield  {journal} {\bibinfo  {journal}
  {Nucl. Phys. A}\ }\textbf {\bibinfo {volume} {896}},\ \bibinfo {pages}
  {46--52} (\bibinfo {year} {2012})},\ \Eprint {http://arxiv.org/abs/1209.6168}
  {arXiv:1209.6168 [nucl-ex]} \BibitemShut {NoStop}%
\bibitem [{\citenamefont {Krasznahorkay}\ \emph {et~al.}(1999)\citenamefont
  {Krasznahorkay} \emph {et~al.}}]{Krasznahorkay:1999zz}%
  \BibitemOpen
  \bibfield  {author} {\bibinfo {author} {\bibfnamefont {A.}~\bibnamefont
  {Krasznahorkay}} \emph {et~al.},\ }\bibfield  {title} {\enquote {\bibinfo
  {title} {{Excitation of Isovector Spin-Dipole Resonances and Neutron Skin of
  Nuclei}},}\ }\href {\doibase 10.1103/PhysRevLett.82.3216} {\bibfield
  {journal} {\bibinfo  {journal} {Phys. Rev. Lett.}\ }\textbf {\bibinfo
  {volume} {82}},\ \bibinfo {pages} {3216--3219} (\bibinfo {year}
  {1999})}\BibitemShut {NoStop}%
\bibitem [{\citenamefont {Tarbert}\ \emph {et~al.}(2014)\citenamefont {Tarbert}
  \emph {et~al.}}]{Tarbert:2013jze}%
  \BibitemOpen
  \bibfield  {author} {\bibinfo {author} {\bibfnamefont {C.~M.}\ \bibnamefont
  {Tarbert}} \emph {et~al.},\ }\bibfield  {title} {\enquote {\bibinfo {title}
  {{Neutron skin of $^{208}$Pb from Coherent Pion Photoproduction}},}\ }\href
  {\doibase 10.1103/PhysRevLett.112.242502} {\bibfield  {journal} {\bibinfo
  {journal} {Phys. Rev. Lett.}\ }\textbf {\bibinfo {volume} {112}},\ \bibinfo
  {pages} {242502} (\bibinfo {year} {2014})},\ \Eprint
  {http://arxiv.org/abs/1311.0168} {arXiv:1311.0168 [nucl-ex]} \BibitemShut
  {NoStop}%
\bibitem [{\citenamefont {Klos}\ \emph {et~al.}(2007)\citenamefont {Klos} \emph
  {et~al.}}]{Klos:2007is}%
  \BibitemOpen
  \bibfield  {author} {\bibinfo {author} {\bibfnamefont {B.}~\bibnamefont
  {Klos}} \emph {et~al.},\ }\bibfield  {title} {\enquote {\bibinfo {title}
  {{Neutron density distributions from antiprotonic Pb-208 and Bi-209
  atoms}},}\ }\href {\doibase 10.1103/PhysRevC.76.014311} {\bibfield  {journal}
  {\bibinfo  {journal} {Phys. Rev. C}\ }\textbf {\bibinfo {volume} {76}},\
  \bibinfo {pages} {014311} (\bibinfo {year} {2007})},\ \Eprint
  {http://arxiv.org/abs/nucl-ex/0702016} {arXiv:nucl-ex/0702016} \BibitemShut
  {NoStop}%
\bibitem [{\citenamefont {Brown}\ \emph {et~al.}(2007)\citenamefont {Brown},
  \citenamefont {Shen}, \citenamefont {Hillhouse}, \citenamefont {Meng},\ and\
  \citenamefont {Trzcinska}}]{Brown:2007zzc}%
  \BibitemOpen
  \bibfield  {author} {\bibinfo {author} {\bibfnamefont {B.~Alex}\ \bibnamefont
  {Brown}}, \bibinfo {author} {\bibfnamefont {G.}~\bibnamefont {Shen}},
  \bibinfo {author} {\bibfnamefont {G.~C.}\ \bibnamefont {Hillhouse}}, \bibinfo
  {author} {\bibfnamefont {Jie}\ \bibnamefont {Meng}}, \ and\ \bibinfo {author}
  {\bibfnamefont {A.}~\bibnamefont {Trzcinska}},\ }\bibfield  {title} {\enquote
  {\bibinfo {title} {{Neutron skin deduced from antiprotonic atom data}},}\
  }\href {\doibase 10.1103/PhysRevC.76.034305} {\bibfield  {journal} {\bibinfo
  {journal} {Phys. Rev. C}\ }\textbf {\bibinfo {volume} {76}},\ \bibinfo
  {pages} {034305} (\bibinfo {year} {2007})}\BibitemShut {NoStop}%
\bibitem [{\citenamefont {Trzcinska}\ \emph {et~al.}(2001)\citenamefont
  {Trzcinska}, \citenamefont {Jastrzebski}, \citenamefont {Lubinski},
  \citenamefont {Hartmann}, \citenamefont {Schmidt}, \citenamefont {von
  Egidy},\ and\ \citenamefont {Klos}}]{Trzcinska:2001sy}%
  \BibitemOpen
  \bibfield  {author} {\bibinfo {author} {\bibfnamefont {A.}~\bibnamefont
  {Trzcinska}}, \bibinfo {author} {\bibfnamefont {J.}~\bibnamefont
  {Jastrzebski}}, \bibinfo {author} {\bibfnamefont {P.}~\bibnamefont
  {Lubinski}}, \bibinfo {author} {\bibfnamefont {F.~J.}\ \bibnamefont
  {Hartmann}}, \bibinfo {author} {\bibfnamefont {R.}~\bibnamefont {Schmidt}},
  \bibinfo {author} {\bibfnamefont {T.}~\bibnamefont {von Egidy}}, \ and\
  \bibinfo {author} {\bibfnamefont {B.}~\bibnamefont {Klos}},\ }\bibfield
  {title} {\enquote {\bibinfo {title} {{Neutron density distributions deduced
  from anti-protonic atoms}},}\ }\href {\doibase 10.1103/PhysRevLett.87.082501}
  {\bibfield  {journal} {\bibinfo  {journal} {Phys. Rev. Lett.}\ }\textbf
  {\bibinfo {volume} {87}},\ \bibinfo {pages} {082501} (\bibinfo {year}
  {2001})}\BibitemShut {NoStop}%
\bibitem [{\citenamefont {Adhikari}\ \emph {et~al.}(2021)\citenamefont
  {Adhikari} \emph {et~al.}}]{PREX:2021umo}%
  \BibitemOpen
  \bibfield  {author} {\bibinfo {author} {\bibfnamefont {D.}~\bibnamefont
  {Adhikari}} \emph {et~al.} (\bibinfo {collaboration} {PREX}),\ }\bibfield
  {title} {\enquote {\bibinfo {title} {{Accurate Determination of the Neutron
  Skin Thickness of $^{208}$Pb through Parity-Violation in Electron
  Scattering}},}\ }\href {\doibase 10.1103/PhysRevLett.126.172502} {\bibfield
  {journal} {\bibinfo  {journal} {Phys. Rev. Lett.}\ }\textbf {\bibinfo
  {volume} {126}},\ \bibinfo {pages} {172502} (\bibinfo {year} {2021})},\
  \Eprint {http://arxiv.org/abs/2102.10767} {arXiv:2102.10767 [nucl-ex]}
  \BibitemShut {NoStop}%
\bibitem [{\citenamefont {Adhikari}\ \emph {et~al.}(2022)\citenamefont
  {Adhikari} \emph {et~al.}}]{CREX:2022kgg}%
  \BibitemOpen
  \bibfield  {author} {\bibinfo {author} {\bibfnamefont {D.}~\bibnamefont
  {Adhikari}} \emph {et~al.} (\bibinfo {collaboration} {CREX}),\ }\bibfield
  {title} {\enquote {\bibinfo {title} {{Precision Determination of the Neutral
  Weak Form Factor of Ca48}},}\ }\href {\doibase
  10.1103/PhysRevLett.129.042501} {\bibfield  {journal} {\bibinfo  {journal}
  {Phys. Rev. Lett.}\ }\textbf {\bibinfo {volume} {129}},\ \bibinfo {pages}
  {042501} (\bibinfo {year} {2022})},\ \Eprint
  {http://arxiv.org/abs/2205.11593} {arXiv:2205.11593 [nucl-ex]} \BibitemShut
  {NoStop}%
\bibitem [{\citenamefont {Li}\ \emph {et~al.}(2020)\citenamefont {Li},
  \citenamefont {Xu}, \citenamefont {Zhou}, \citenamefont {Wang}, \citenamefont
  {Zhao}, \citenamefont {Chen},\ and\ \citenamefont {Wang}}]{Li:2019kkh}%
  \BibitemOpen
  \bibfield  {author} {\bibinfo {author} {\bibfnamefont {Hanlin}\ \bibnamefont
  {Li}}, \bibinfo {author} {\bibfnamefont {Hao-jie}\ \bibnamefont {Xu}},
  \bibinfo {author} {\bibfnamefont {Ying}\ \bibnamefont {Zhou}}, \bibinfo
  {author} {\bibfnamefont {Xiaobao}\ \bibnamefont {Wang}}, \bibinfo {author}
  {\bibfnamefont {Jie}\ \bibnamefont {Zhao}}, \bibinfo {author} {\bibfnamefont
  {Lie-Wen}\ \bibnamefont {Chen}}, \ and\ \bibinfo {author} {\bibfnamefont
  {Fuqiang}\ \bibnamefont {Wang}},\ }\bibfield  {title} {\enquote {\bibinfo
  {title} {{Probing the neutron skin with ultrarelativistic isobaric
  collisions}},}\ }\href {\doibase 10.1103/PhysRevLett.125.222301} {\bibfield
  {journal} {\bibinfo  {journal} {Phys. Rev. Lett.}\ }\textbf {\bibinfo
  {volume} {125}},\ \bibinfo {pages} {222301} (\bibinfo {year} {2020})},\
  \Eprint {http://arxiv.org/abs/1910.06170} {arXiv:1910.06170 [nucl-th]}
  \BibitemShut {NoStop}%
\bibitem [{\citenamefont {Giacalone}\ \emph {et~al.}(2023)\citenamefont
  {Giacalone}, \citenamefont {Nijs},\ and\ \citenamefont {van~der
  Schee}}]{Giacalone:2023cet}%
  \BibitemOpen
  \bibfield  {author} {\bibinfo {author} {\bibfnamefont {Giuliano}\
  \bibnamefont {Giacalone}}, \bibinfo {author} {\bibfnamefont {Govert}\
  \bibnamefont {Nijs}}, \ and\ \bibinfo {author} {\bibfnamefont {Wilke}\
  \bibnamefont {van~der Schee}},\ }\bibfield  {title} {\enquote {\bibinfo
  {title} {{Determination of the Neutron Skin of Pb208 from Ultrarelativistic
  Nuclear Collisions}},}\ }\href {\doibase 10.1103/PhysRevLett.131.202302}
  {\bibfield  {journal} {\bibinfo  {journal} {Phys. Rev. Lett.}\ }\textbf
  {\bibinfo {volume} {131}},\ \bibinfo {pages} {202302} (\bibinfo {year}
  {2023})},\ \Eprint {http://arxiv.org/abs/2305.00015} {arXiv:2305.00015
  [nucl-th]} \BibitemShut {NoStop}%
\bibitem [{\citenamefont {Liu}\ \emph {et~al.}(2022{\natexlab{a}})\citenamefont
  {Liu}, \citenamefont {Zhang}, \citenamefont {Zhou}, \citenamefont {Xu},
  \citenamefont {Jia},\ and\ \citenamefont {Peng}}]{Liu:2022kvz}%
  \BibitemOpen
  \bibfield  {author} {\bibinfo {author} {\bibfnamefont {Lu-Meng}\ \bibnamefont
  {Liu}}, \bibinfo {author} {\bibfnamefont {Chun-Jian}\ \bibnamefont {Zhang}},
  \bibinfo {author} {\bibfnamefont {Jia}\ \bibnamefont {Zhou}}, \bibinfo
  {author} {\bibfnamefont {Jun}\ \bibnamefont {Xu}}, \bibinfo {author}
  {\bibfnamefont {Jiangyong}\ \bibnamefont {Jia}}, \ and\ \bibinfo {author}
  {\bibfnamefont {Guang-Xiong}\ \bibnamefont {Peng}},\ }\bibfield  {title}
  {\enquote {\bibinfo {title} {{Probing neutron-skin thickness with free
  spectator neutrons in ultracentral high-energy isobaric collisions}},}\
  }\href {\doibase 10.1016/j.physletb.2022.137441} {\bibfield  {journal}
  {\bibinfo  {journal} {Phys. Lett. B}\ }\textbf {\bibinfo {volume} {834}},\
  \bibinfo {pages} {137441} (\bibinfo {year} {2022}{\natexlab{a}})},\ \Eprint
  {http://arxiv.org/abs/2203.09924} {arXiv:2203.09924 [nucl-th]} \BibitemShut
  {NoStop}%
\bibitem [{\citenamefont {Liu}\ \emph {et~al.}(2022{\natexlab{b}})\citenamefont
  {Liu}, \citenamefont {Zhang}, \citenamefont {Xu}, \citenamefont {Jia},\ and\
  \citenamefont {Peng}}]{Liu:2022xlm}%
  \BibitemOpen
  \bibfield  {author} {\bibinfo {author} {\bibfnamefont {Lu-Meng}\ \bibnamefont
  {Liu}}, \bibinfo {author} {\bibfnamefont {Chun-Jian}\ \bibnamefont {Zhang}},
  \bibinfo {author} {\bibfnamefont {Jun}\ \bibnamefont {Xu}}, \bibinfo {author}
  {\bibfnamefont {Jiangyong}\ \bibnamefont {Jia}}, \ and\ \bibinfo {author}
  {\bibfnamefont {Guang-Xiong}\ \bibnamefont {Peng}},\ }\bibfield  {title}
  {\enquote {\bibinfo {title} {{Free spectator nucleons in ultracentral
  relativistic heavy-ion collisions as a probe of neutron skin}},}\ }\href
  {\doibase 10.1103/PhysRevC.106.034913} {\bibfield  {journal} {\bibinfo
  {journal} {Phys. Rev. C}\ }\textbf {\bibinfo {volume} {106}},\ \bibinfo
  {pages} {034913} (\bibinfo {year} {2022}{\natexlab{b}})},\ \Eprint
  {http://arxiv.org/abs/2209.03106} {arXiv:2209.03106 [nucl-th]} \BibitemShut
  {NoStop}%
\bibitem [{\citenamefont {Liu}\ \emph {et~al.}(2023{\natexlab{a}})\citenamefont
  {Liu}, \citenamefont {Xu},\ and\ \citenamefont {Peng}}]{Liu:2023qeq}%
  \BibitemOpen
  \bibfield  {author} {\bibinfo {author} {\bibfnamefont {Lu-Meng}\ \bibnamefont
  {Liu}}, \bibinfo {author} {\bibfnamefont {Jun}\ \bibnamefont {Xu}}, \ and\
  \bibinfo {author} {\bibfnamefont {Guang-Xiong}\ \bibnamefont {Peng}},\
  }\bibfield  {title} {\enquote {\bibinfo {title} {{Measuring deformed neutron
  skin with free spectator nucleons in relativistic heavy-ion collisions}},}\
  }\href {\doibase 10.1016/j.physletb.2023.137701} {\bibfield  {journal}
  {\bibinfo  {journal} {Phys. Lett. B}\ }\textbf {\bibinfo {volume} {838}},\
  \bibinfo {pages} {137701} (\bibinfo {year} {2023}{\natexlab{a}})},\ \Eprint
  {http://arxiv.org/abs/2301.07893} {arXiv:2301.07893 [nucl-th]} \BibitemShut
  {NoStop}%
\bibitem [{\citenamefont {Liu}\ \emph {et~al.}(2023{\natexlab{b}})\citenamefont
  {Liu}, \citenamefont {Xu},\ and\ \citenamefont {Peng}}]{Liu:2023rap}%
  \BibitemOpen
  \bibfield  {author} {\bibinfo {author} {\bibfnamefont {Lu-Meng}\ \bibnamefont
  {Liu}}, \bibinfo {author} {\bibfnamefont {Jun}\ \bibnamefont {Xu}}, \ and\
  \bibinfo {author} {\bibfnamefont {Guang-Xiong}\ \bibnamefont {Peng}},\
  }\bibfield  {title} {\enquote {\bibinfo {title} {{Collision geometry effect
  on free spectator nucleons in relativistic heavy-ion collisions}},}\ }\href
  {\doibase 10.11804/NuclPhysRev.40.2022095} {\bibfield  {journal} {\bibinfo
  {journal} {Nucl. Phys. Rev.}\ }\textbf {\bibinfo {volume} {40}},\ \bibinfo
  {pages} {2022095} (\bibinfo {year} {2023}{\natexlab{b}})},\ \Eprint
  {http://arxiv.org/abs/2301.08251} {arXiv:2301.08251 [nucl-th]} \BibitemShut
  {NoStop}%
\bibitem [{\citenamefont {Sun}\ \emph {et~al.}(2010)\citenamefont {Sun},
  \citenamefont {Fang}, \citenamefont {Ma}, \citenamefont {Cai}, \citenamefont
  {Chen}, \citenamefont {Guo}, \citenamefont {Tian},\ and\ \citenamefont
  {Wang}}]{Sun:2009wf}%
  \BibitemOpen
  \bibfield  {author} {\bibinfo {author} {\bibfnamefont {X.~Y.}\ \bibnamefont
  {Sun}}, \bibinfo {author} {\bibfnamefont {D.~Q.}\ \bibnamefont {Fang}},
  \bibinfo {author} {\bibfnamefont {Y.~G.}\ \bibnamefont {Ma}}, \bibinfo
  {author} {\bibfnamefont {X.~Z.}\ \bibnamefont {Cai}}, \bibinfo {author}
  {\bibfnamefont {J.~G.}\ \bibnamefont {Chen}}, \bibinfo {author}
  {\bibfnamefont {W.}~\bibnamefont {Guo}}, \bibinfo {author} {\bibfnamefont
  {W.~D.}\ \bibnamefont {Tian}}, \ and\ \bibinfo {author} {\bibfnamefont
  {H.~W.}\ \bibnamefont {Wang}},\ }\bibfield  {title} {\enquote {\bibinfo
  {title} {{Neutron/proton ratio of nucleon emissions as a probe of neutron
  skin}},}\ }\href {\doibase 10.1016/j.physletb.2009.11.031} {\bibfield
  {journal} {\bibinfo  {journal} {Phys. Lett. B}\ }\textbf {\bibinfo {volume}
  {682}},\ \bibinfo {pages} {396--400} (\bibinfo {year} {2010})},\ \Eprint
  {http://arxiv.org/abs/0906.5281} {arXiv:0906.5281 [nucl-th]} \BibitemShut
  {NoStop}%
\bibitem [{\citenamefont {Dai}\ \emph {et~al.}(2014)\citenamefont {Dai},
  \citenamefont {Fang}, \citenamefont {Ma}, \citenamefont {Cao},\ and\
  \citenamefont {Zhang}}]{Dai:2014rja}%
  \BibitemOpen
  \bibfield  {author} {\bibinfo {author} {\bibfnamefont {Z.~T.}\ \bibnamefont
  {Dai}}, \bibinfo {author} {\bibfnamefont {D.~Q.}\ \bibnamefont {Fang}},
  \bibinfo {author} {\bibfnamefont {Y.~G.}\ \bibnamefont {Ma}}, \bibinfo
  {author} {\bibfnamefont {X.~G.}\ \bibnamefont {Cao}}, \ and\ \bibinfo
  {author} {\bibfnamefont {G.~Q.}\ \bibnamefont {Zhang}},\ }\bibfield  {title}
  {\enquote {\bibinfo {title} {{Triton/$^{3}$He ratio as an observable for
  neutron skin thickness}},}\ }\href {\doibase 10.1103/PhysRevC.89.014613}
  {\bibfield  {journal} {\bibinfo  {journal} {Phys. Rev. C}\ }\textbf {\bibinfo
  {volume} {89}},\ \bibinfo {pages} {014613} (\bibinfo {year} {2014})},\
  \Eprint {http://arxiv.org/abs/1402.3038} {arXiv:1402.3038 [nucl-th]}
  \BibitemShut {NoStop}%
\bibitem [{\citenamefont {Dai}\ \emph {et~al.}(2015)\citenamefont {Dai},
  \citenamefont {Fang}, \citenamefont {Ma}, \citenamefont {Cao}, \citenamefont
  {Zhang},\ and\ \citenamefont {Shen}}]{Dai:2015dua}%
  \BibitemOpen
  \bibfield  {author} {\bibinfo {author} {\bibfnamefont {Z.~T.}\ \bibnamefont
  {Dai}}, \bibinfo {author} {\bibfnamefont {D.~Q.}\ \bibnamefont {Fang}},
  \bibinfo {author} {\bibfnamefont {Y.~G.}\ \bibnamefont {Ma}}, \bibinfo
  {author} {\bibfnamefont {X.~G.}\ \bibnamefont {Cao}}, \bibinfo {author}
  {\bibfnamefont {G.~Q.}\ \bibnamefont {Zhang}}, \ and\ \bibinfo {author}
  {\bibfnamefont {W.~Q.}\ \bibnamefont {Shen}},\ }\bibfield  {title} {\enquote
  {\bibinfo {title} {{Effect of neutron skin thickness on projectile
  fragmentation}},}\ }\href {\doibase 10.1103/PhysRevC.91.034618} {\bibfield
  {journal} {\bibinfo  {journal} {Phys. Rev. C}\ }\textbf {\bibinfo {volume}
  {91}},\ \bibinfo {pages} {034618} (\bibinfo {year} {2015})},\ \Eprint
  {http://arxiv.org/abs/1504.00226} {arXiv:1504.00226 [nucl-th]} \BibitemShut
  {NoStop}%
\bibitem [{\citenamefont {Ding}\ \emph {et~al.}(2024)\citenamefont {Ding},
  \citenamefont {Fang},\ and\ \citenamefont {Ma}}]{Ding:2024jfr}%
  \BibitemOpen
  \bibfield  {author} {\bibinfo {author} {\bibfnamefont {Meng-Qi}\ \bibnamefont
  {Ding}}, \bibinfo {author} {\bibfnamefont {De-Qing}\ \bibnamefont {Fang}}, \
  and\ \bibinfo {author} {\bibfnamefont {Yu-Gang}\ \bibnamefont {Ma}},\
  }\bibfield  {title} {\enquote {\bibinfo {title} {{Effects of neutron-skin
  thickness on light-particle production}},}\ }\href {\doibase
  10.1103/PhysRevC.109.024616} {\bibfield  {journal} {\bibinfo  {journal}
  {Phys. Rev. C}\ }\textbf {\bibinfo {volume} {109}},\ \bibinfo {pages}
  {024616} (\bibinfo {year} {2024})}\BibitemShut {NoStop}%
\bibitem [{\citenamefont {Yang}\ \emph {et~al.}(2023)\citenamefont {Yang},
  \citenamefont {Chen}, \citenamefont {Cui}, \citenamefont {Li},\ and\
  \citenamefont {Zhang}}]{Yang:2023bwm}%
  \BibitemOpen
  \bibfield  {author} {\bibinfo {author} {\bibfnamefont {Junping}\ \bibnamefont
  {Yang}}, \bibinfo {author} {\bibfnamefont {Xiang}\ \bibnamefont {Chen}},
  \bibinfo {author} {\bibfnamefont {Ying}\ \bibnamefont {Cui}}, \bibinfo
  {author} {\bibfnamefont {Zhuxia}\ \bibnamefont {Li}}, \ and\ \bibinfo
  {author} {\bibfnamefont {Yingxun}\ \bibnamefont {Zhang}},\ }\bibfield
  {title} {\enquote {\bibinfo {title} {{Probing the Neutron Skin of Unstable
  Nuclei with Heavy-Ion Collisions}},}\ }\href {\doibase
  10.3390/universe9050206} {\bibfield  {journal} {\bibinfo  {journal}
  {Universe}\ }\textbf {\bibinfo {volume} {9}},\ \bibinfo {pages} {206}
  (\bibinfo {year} {2023})},\ \Eprint {http://arxiv.org/abs/2304.12059}
  {arXiv:2304.12059 [nucl-th]} \BibitemShut {NoStop}%
\bibitem [{\citenamefont {Upadhyaya}\ \emph {et~al.}(2024)\citenamefont
  {Upadhyaya} \emph {et~al.}}]{FAZIA:2024vfo}%
  \BibitemOpen
  \bibfield  {author} {\bibinfo {author} {\bibfnamefont {S.}~\bibnamefont
  {Upadhyaya}} \emph {et~al.} (\bibinfo {collaboration} {FAZIA}),\ }\bibfield
  {title} {\enquote {\bibinfo {title} {{Study of quasi-projectile properties at
  Fermi energies in $^{48}$Ca projectile systems}},}\ }\href@noop {} {\
  (\bibinfo {year} {2024})},\ \Eprint {http://arxiv.org/abs/2402.09289}
  {arXiv:2402.09289 [nucl-ex]} \BibitemShut {NoStop}%
\bibitem [{\citenamefont {Hagel}\ \emph {et~al.}(1994)\citenamefont {Hagel}
  \emph {et~al.}}]{Hagel:1994zz}%
  \BibitemOpen
  \bibfield  {author} {\bibinfo {author} {\bibfnamefont {K.}~\bibnamefont
  {Hagel}} \emph {et~al.},\ }\bibfield  {title} {\enquote {\bibinfo {title}
  {{Violent collisions and multifragment final states in the Ca-40 + Ca-40
  reaction at 35 MeV/nucleon}},}\ }\href {\doibase 10.1103/PhysRevC.50.2017}
  {\bibfield  {journal} {\bibinfo  {journal} {Phys. Rev. C}\ }\textbf {\bibinfo
  {volume} {50}},\ \bibinfo {pages} {2017--2034} (\bibinfo {year}
  {1994})}\BibitemShut {NoStop}%
\bibitem [{\citenamefont {Mocko}\ \emph {et~al.}(2006)\citenamefont {Mocko}
  \emph {et~al.}}]{Mocko:2006tt}%
  \BibitemOpen
  \bibfield  {author} {\bibinfo {author} {\bibfnamefont {M.}~\bibnamefont
  {Mocko}} \emph {et~al.},\ }\bibfield  {title} {\enquote {\bibinfo {title}
  {{Projectile fragmentation of Ca40, Ca48, Ni58, and Ni64 at 140
  MeV/nucleon}},}\ }\href {\doibase 10.1103/PhysRevC.74.054612} {\bibfield
  {journal} {\bibinfo  {journal} {Phys. Rev. C}\ }\textbf {\bibinfo {volume}
  {74}},\ \bibinfo {pages} {054612} (\bibinfo {year} {2006})}\BibitemShut
  {NoStop}%
\bibitem [{\citenamefont {Pawlowski}\ \emph {et~al.}(2023)\citenamefont
  {Pawlowski} \emph {et~al.}}]{Pawlowski:2023gen}%
  \BibitemOpen
  \bibfield  {author} {\bibinfo {author} {\bibfnamefont {P.}~\bibnamefont
  {Pawlowski}} \emph {et~al.},\ }\bibfield  {title} {\enquote {\bibinfo {title}
  {{Neutrons from projectile fragmentation at 600 MeV/nucleon}},}\ }\href
  {\doibase 10.1103/PhysRevC.108.044610} {\bibfield  {journal} {\bibinfo
  {journal} {Phys. Rev. C}\ }\textbf {\bibinfo {volume} {108}},\ \bibinfo
  {pages} {044610} (\bibinfo {year} {2023})},\ \Eprint
  {http://arxiv.org/abs/2310.00409} {arXiv:2310.00409 [nucl-ex]} \BibitemShut
  {NoStop}%
\bibitem [{\citenamefont {Ma}\ \emph {et~al.}(2021)\citenamefont {Ma},
  \citenamefont {Wei}, \citenamefont {Liu}, \citenamefont {Su}, \citenamefont
  {Zheng}, \citenamefont {Lin},\ and\ \citenamefont {Zhang}}]{Ma:2021nwr}%
  \BibitemOpen
  \bibfield  {author} {\bibinfo {author} {\bibfnamefont {Chun-Wang}\
  \bibnamefont {Ma}}, \bibinfo {author} {\bibfnamefont {Hui-Ling}\ \bibnamefont
  {Wei}}, \bibinfo {author} {\bibfnamefont {Xing-Quan}\ \bibnamefont {Liu}},
  \bibinfo {author} {\bibfnamefont {Jun}\ \bibnamefont {Su}}, \bibinfo {author}
  {\bibfnamefont {Hua}\ \bibnamefont {Zheng}}, \bibinfo {author} {\bibfnamefont
  {Wei-Ping}\ \bibnamefont {Lin}}, \ and\ \bibinfo {author} {\bibfnamefont
  {Ying-Xun}\ \bibnamefont {Zhang}},\ }\bibfield  {title} {\enquote {\bibinfo
  {title} {{Nuclear fragments in projectile fragmentation reactions}},}\ }\href
  {\doibase 10.1016/j.ppnp.2021.103911} {\bibfield  {journal} {\bibinfo
  {journal} {Prog. Part. Nucl. Phys.}\ }\textbf {\bibinfo {volume} {121}},\
  \bibinfo {pages} {103911} (\bibinfo {year} {2021})}\BibitemShut {NoStop}%
\bibitem [{\citenamefont {Aumann}\ \emph {et~al.}(2017)\citenamefont {Aumann},
  \citenamefont {Bertulani}, \citenamefont {Schindler},\ and\ \citenamefont
  {Typel}}]{PhysRevLett.119.262501}%
  \BibitemOpen
  \bibfield  {author} {\bibinfo {author} {\bibfnamefont {T.}~\bibnamefont
  {Aumann}}, \bibinfo {author} {\bibfnamefont {C.~A.}\ \bibnamefont
  {Bertulani}}, \bibinfo {author} {\bibfnamefont {F.}~\bibnamefont
  {Schindler}}, \ and\ \bibinfo {author} {\bibfnamefont {S.}~\bibnamefont
  {Typel}},\ }\bibfield  {title} {\enquote {\bibinfo {title} {Peeling off
  neutron skins from neutron-rich nuclei: Constraints on the symmetry energy
  from neutron-removal cross sections},}\ }\href {\doibase
  10.1103/PhysRevLett.119.262501} {\bibfield  {journal} {\bibinfo  {journal}
  {Phys. Rev. Lett.}\ }\textbf {\bibinfo {volume} {119}},\ \bibinfo {pages}
  {262501} (\bibinfo {year} {2017})}\BibitemShut {NoStop}%
\bibitem [{\citenamefont {Wei}\ \emph {et~al.}(2014)\citenamefont {Wei},
  \citenamefont {Li}, \citenamefont {Xu},\ and\ \citenamefont
  {Chen}}]{Wei:2013sfa}%
  \BibitemOpen
  \bibfield  {author} {\bibinfo {author} {\bibfnamefont {Gao-Feng}\
  \bibnamefont {Wei}}, \bibinfo {author} {\bibfnamefont {Bao-An}\ \bibnamefont
  {Li}}, \bibinfo {author} {\bibfnamefont {Jun}\ \bibnamefont {Xu}}, \ and\
  \bibinfo {author} {\bibfnamefont {Lie-Wen}\ \bibnamefont {Chen}},\ }\bibfield
   {title} {\enquote {\bibinfo {title} {{Influence of neutron-skin thickness on
  $\pi^{-}/\pi^{+}$ ratio in Pb+Pb collisions}},}\ }\href {\doibase
  10.1103/PhysRevC.90.014610} {\bibfield  {journal} {\bibinfo  {journal} {Phys.
  Rev. C}\ }\textbf {\bibinfo {volume} {90}},\ \bibinfo {pages} {014610}
  (\bibinfo {year} {2014})},\ \Eprint {http://arxiv.org/abs/1309.7717}
  {arXiv:1309.7717 [nucl-th]} \BibitemShut {NoStop}%
\bibitem [{\citenamefont {Hartnack}\ \emph {et~al.}(2018)\citenamefont
  {Hartnack}, \citenamefont {Le~F\`evre}, \citenamefont {Leifels},\ and\
  \citenamefont {Aichelin}}]{Hartnack:2018sih}%
  \BibitemOpen
  \bibfield  {author} {\bibinfo {author} {\bibfnamefont {C.}~\bibnamefont
  {Hartnack}}, \bibinfo {author} {\bibfnamefont {A.}~\bibnamefont
  {Le~F\`evre}}, \bibinfo {author} {\bibfnamefont {Y.}~\bibnamefont {Leifels}},
  \ and\ \bibinfo {author} {\bibfnamefont {J.}~\bibnamefont {Aichelin}},\
  }\bibfield  {title} {\enquote {\bibinfo {title} {{The influence of the
  neutron skin and the asymmetry energy on the $\pi^-/\pi^+$ ratio}},}\
  }\href@noop {} {\  (\bibinfo {year} {2018})},\ \Eprint
  {http://arxiv.org/abs/1808.09868} {arXiv:1808.09868 [nucl-th]} \BibitemShut
  {NoStop}%
\bibitem [{\citenamefont {Li}\ \emph {et~al.}(1997)\citenamefont {Li},
  \citenamefont {Ko},\ and\ \citenamefont {Ren}}]{Li:1997rc}%
  \BibitemOpen
  \bibfield  {author} {\bibinfo {author} {\bibfnamefont {Bao-An}\ \bibnamefont
  {Li}}, \bibinfo {author} {\bibfnamefont {C.~M.}\ \bibnamefont {Ko}}, \ and\
  \bibinfo {author} {\bibfnamefont {Zhong-zhou}\ \bibnamefont {Ren}},\
  }\bibfield  {title} {\enquote {\bibinfo {title} {{Equation of state of
  asymmetric nuclear matter and collisions of neutron rich nuclei}},}\ }\href
  {\doibase 10.1103/PhysRevLett.78.1644} {\bibfield  {journal} {\bibinfo
  {journal} {Phys. Rev. Lett.}\ }\textbf {\bibinfo {volume} {78}},\ \bibinfo
  {pages} {1644} (\bibinfo {year} {1997})},\ \Eprint
  {http://arxiv.org/abs/nucl-th/9701048} {arXiv:nucl-th/9701048} \BibitemShut
  {NoStop}%
\bibitem [{\citenamefont {Zhang}\ \emph {et~al.}(2008)\citenamefont {Zhang},
  \citenamefont {Danielewicz}, \citenamefont {Famiano}, \citenamefont {Li},
  \citenamefont {Lynch}, \citenamefont {Tsang},\ and\ \citenamefont
  {Li}}]{Zhang:2007hmv}%
  \BibitemOpen
  \bibfield  {author} {\bibinfo {author} {\bibfnamefont {Y.~X.}\ \bibnamefont
  {Zhang}}, \bibinfo {author} {\bibfnamefont {P.}~\bibnamefont {Danielewicz}},
  \bibinfo {author} {\bibfnamefont {M.}~\bibnamefont {Famiano}}, \bibinfo
  {author} {\bibfnamefont {Z.}~\bibnamefont {Li}}, \bibinfo {author}
  {\bibfnamefont {W.~G.}\ \bibnamefont {Lynch}}, \bibinfo {author}
  {\bibfnamefont {M.~B.}\ \bibnamefont {Tsang}}, \ and\ \bibinfo {author}
  {\bibfnamefont {Zhuxia}\ \bibnamefont {Li}},\ }\bibfield  {title} {\enquote
  {\bibinfo {title} {{The influence of cluster emission and the symmetry energy
  on neutron-proton spectral double ratios}},}\ }\href {\doibase
  10.1016/j.physletb.2008.03.075} {\bibfield  {journal} {\bibinfo  {journal}
  {Phys. Lett. B}\ }\textbf {\bibinfo {volume} {664}},\ \bibinfo {pages}
  {145--148} (\bibinfo {year} {2008})},\ \Eprint
  {http://arxiv.org/abs/0708.3684} {arXiv:0708.3684 [nucl-th]} \BibitemShut
  {NoStop}%
\bibitem [{\citenamefont {Li}\ \emph {et~al.}(2006)\citenamefont {Li},
  \citenamefont {Chen}, \citenamefont {Yong},\ and\ \citenamefont
  {Zuo}}]{Li:2005by}%
  \BibitemOpen
  \bibfield  {author} {\bibinfo {author} {\bibfnamefont {Bao-An}\ \bibnamefont
  {Li}}, \bibinfo {author} {\bibfnamefont {Lie-Wen}\ \bibnamefont {Chen}},
  \bibinfo {author} {\bibfnamefont {Gao-Chan}\ \bibnamefont {Yong}}, \ and\
  \bibinfo {author} {\bibfnamefont {Wei}\ \bibnamefont {Zuo}},\ }\bibfield
  {title} {\enquote {\bibinfo {title} {{Double neutron/proton ratio of nucleon
  emissions in isotopic reaction systems as a robust probe of nuclear symmetry
  energy}},}\ }\href {\doibase 10.1016/j.physletb.2006.02.003} {\bibfield
  {journal} {\bibinfo  {journal} {Phys. Lett. B}\ }\textbf {\bibinfo {volume}
  {634}},\ \bibinfo {pages} {378--382} (\bibinfo {year} {2006})},\ \Eprint
  {http://arxiv.org/abs/nucl-th/0510016} {arXiv:nucl-th/0510016} \BibitemShut
  {NoStop}%
\bibitem [{\citenamefont {Famiano}\ \emph {et~al.}(2006)\citenamefont
  {Famiano}, \citenamefont {Liu}, \citenamefont {Lynch}, \citenamefont
  {Rogers}, \citenamefont {Tsang}, \citenamefont {Wallace}, \citenamefont
  {Charity}, \citenamefont {Komarov}, \citenamefont {Sarantites},\ and\
  \citenamefont {Sobotka}}]{Famiano:2006rb}%
  \BibitemOpen
  \bibfield  {author} {\bibinfo {author} {\bibfnamefont {M.~A.}\ \bibnamefont
  {Famiano}}, \bibinfo {author} {\bibfnamefont {T.}~\bibnamefont {Liu}},
  \bibinfo {author} {\bibfnamefont {W.~G.}\ \bibnamefont {Lynch}}, \bibinfo
  {author} {\bibfnamefont {A.~M.}\ \bibnamefont {Rogers}}, \bibinfo {author}
  {\bibfnamefont {M.~B.}\ \bibnamefont {Tsang}}, \bibinfo {author}
  {\bibfnamefont {M.~S.}\ \bibnamefont {Wallace}}, \bibinfo {author}
  {\bibfnamefont {R.~J.}\ \bibnamefont {Charity}}, \bibinfo {author}
  {\bibfnamefont {S.}~\bibnamefont {Komarov}}, \bibinfo {author} {\bibfnamefont
  {D.~G.}\ \bibnamefont {Sarantites}}, \ and\ \bibinfo {author} {\bibfnamefont
  {L.~G.}\ \bibnamefont {Sobotka}},\ }\bibfield  {title} {\enquote {\bibinfo
  {title} {{Neutron and Proton Transverse Emission Ratio Measurements and the
  Density Dependence of the Asymmetry Term of the Nuclear Equation of
  State}},}\ }\href {\doibase 10.1103/PhysRevLett.97.052701} {\bibfield
  {journal} {\bibinfo  {journal} {Phys. Rev. Lett.}\ }\textbf {\bibinfo
  {volume} {97}},\ \bibinfo {pages} {052701} (\bibinfo {year} {2006})},\
  \Eprint {http://arxiv.org/abs/nucl-ex/0607016} {arXiv:nucl-ex/0607016}
  \BibitemShut {NoStop}%
\bibitem [{\citenamefont {Rizzo}\ \emph {et~al.}(2005)\citenamefont {Rizzo},
  \citenamefont {Colonna},\ and\ \citenamefont {Di~Toro}}]{Rizzo:2005mk}%
  \BibitemOpen
  \bibfield  {author} {\bibinfo {author} {\bibfnamefont {J.}~\bibnamefont
  {Rizzo}}, \bibinfo {author} {\bibfnamefont {M.}~\bibnamefont {Colonna}}, \
  and\ \bibinfo {author} {\bibfnamefont {M.}~\bibnamefont {Di~Toro}},\
  }\bibfield  {title} {\enquote {\bibinfo {title} {{Fast nucleon emission as a
  probe of the isospin momentum dependence}},}\ }\href {\doibase
  10.1103/PhysRevC.72.064609} {\bibfield  {journal} {\bibinfo  {journal} {Phys.
  Rev. C}\ }\textbf {\bibinfo {volume} {72}},\ \bibinfo {pages} {064609}
  (\bibinfo {year} {2005})},\ \Eprint {http://arxiv.org/abs/nucl-th/0508008}
  {arXiv:nucl-th/0508008} \BibitemShut {NoStop}%
\bibitem [{\citenamefont {Kong}\ \emph {et~al.}(2015)\citenamefont {Kong},
  \citenamefont {Xia}, \citenamefont {Xu}, \citenamefont {Chen}, \citenamefont
  {Li},\ and\ \citenamefont {Ma}}]{Kong:2015rla}%
  \BibitemOpen
  \bibfield  {author} {\bibinfo {author} {\bibfnamefont {Hai-Yun}\ \bibnamefont
  {Kong}}, \bibinfo {author} {\bibfnamefont {Yin}\ \bibnamefont {Xia}},
  \bibinfo {author} {\bibfnamefont {Jun}\ \bibnamefont {Xu}}, \bibinfo {author}
  {\bibfnamefont {Lie-Wen}\ \bibnamefont {Chen}}, \bibinfo {author}
  {\bibfnamefont {Bao-An}\ \bibnamefont {Li}}, \ and\ \bibinfo {author}
  {\bibfnamefont {Yu-Gang}\ \bibnamefont {Ma}},\ }\bibfield  {title} {\enquote
  {\bibinfo {title} {{Reexamination of the neutron-to-proton-ratio puzzle in
  intermediate-energy heavy-ion collisions}},}\ }\href {\doibase
  10.1103/PhysRevC.91.047601} {\bibfield  {journal} {\bibinfo  {journal} {Phys.
  Rev. C}\ }\textbf {\bibinfo {volume} {91}},\ \bibinfo {pages} {047601}
  (\bibinfo {year} {2015})},\ \Eprint {http://arxiv.org/abs/1502.00778}
  {arXiv:1502.00778 [nucl-th]} \BibitemShut {NoStop}%
\bibitem [{\citenamefont {Coupland}\ \emph {et~al.}(2016)\citenamefont
  {Coupland} \emph {et~al.}}]{Coupland:2014gya}%
  \BibitemOpen
  \bibfield  {author} {\bibinfo {author} {\bibfnamefont {D.~D.~S.}\
  \bibnamefont {Coupland}} \emph {et~al.},\ }\bibfield  {title} {\enquote
  {\bibinfo {title} {{Probing effective nucleon masses with heavy-ion
  collisions}},}\ }\href {\doibase 10.1103/PhysRevC.94.011601} {\bibfield
  {journal} {\bibinfo  {journal} {Phys. Rev. C}\ }\textbf {\bibinfo {volume}
  {94}},\ \bibinfo {pages} {011601} (\bibinfo {year} {2016})},\ \Eprint
  {http://arxiv.org/abs/1406.4546} {arXiv:1406.4546 [nucl-ex]} \BibitemShut
  {NoStop}%
\bibitem [{\citenamefont {Morfouace}\ \emph {et~al.}(2019)\citenamefont
  {Morfouace} \emph {et~al.}}]{Morfouace:2019jky}%
  \BibitemOpen
  \bibfield  {author} {\bibinfo {author} {\bibfnamefont {P.}~\bibnamefont
  {Morfouace}} \emph {et~al.},\ }\bibfield  {title} {\enquote {\bibinfo {title}
  {{Constraining the symmetry energy with heavy-ion collisions and Bayesian
  analyses}},}\ }\href {\doibase 10.1016/j.physletb.2019.135045} {\bibfield
  {journal} {\bibinfo  {journal} {Phys. Lett. B}\ }\textbf {\bibinfo {volume}
  {799}},\ \bibinfo {pages} {135045} (\bibinfo {year} {2019})},\ \Eprint
  {http://arxiv.org/abs/1904.12471} {arXiv:1904.12471 [nucl-ex]} \BibitemShut
  {NoStop}%
\bibitem [{\citenamefont {Xu}\ \emph {et~al.}(2015)\citenamefont {Xu},
  \citenamefont {Chen},\ and\ \citenamefont {Li}}]{Xu:2014cwa}%
  \BibitemOpen
  \bibfield  {author} {\bibinfo {author} {\bibfnamefont {Jun}\ \bibnamefont
  {Xu}}, \bibinfo {author} {\bibfnamefont {Lie-Wen}\ \bibnamefont {Chen}}, \
  and\ \bibinfo {author} {\bibfnamefont {Bao-An}\ \bibnamefont {Li}},\
  }\bibfield  {title} {\enquote {\bibinfo {title} {{Thermal properties of
  asymmetric nuclear matter with an improved isospin- and momentum-dependent
  interaction}},}\ }\href {\doibase 10.1103/PhysRevC.91.014611} {\bibfield
  {journal} {\bibinfo  {journal} {Phys. Rev. C}\ }\textbf {\bibinfo {volume}
  {91}},\ \bibinfo {pages} {014611} (\bibinfo {year} {2015})},\ \Eprint
  {http://arxiv.org/abs/1410.1604} {arXiv:1410.1604 [nucl-th]} \BibitemShut
  {NoStop}%
\bibitem [{\citenamefont {Das}\ \emph {et~al.}(2003)\citenamefont {Das},
  \citenamefont {Gupta}, \citenamefont {Gale},\ and\ \citenamefont
  {Li}}]{Das:2002fr}%
  \BibitemOpen
  \bibfield  {author} {\bibinfo {author} {\bibfnamefont {C.~B.}\ \bibnamefont
  {Das}}, \bibinfo {author} {\bibfnamefont {S.~Das}\ \bibnamefont {Gupta}},
  \bibinfo {author} {\bibfnamefont {Charles}\ \bibnamefont {Gale}}, \ and\
  \bibinfo {author} {\bibfnamefont {Bao-An}\ \bibnamefont {Li}},\ }\bibfield
  {title} {\enquote {\bibinfo {title} {{Momentum dependence of symmetry
  potential in asymmetric nuclear matter for transport model calculations}},}\
  }\href {\doibase 10.1103/PhysRevC.67.034611} {\bibfield  {journal} {\bibinfo
  {journal} {Phys. Rev. C}\ }\textbf {\bibinfo {volume} {67}},\ \bibinfo
  {pages} {034611} (\bibinfo {year} {2003})},\ \Eprint
  {http://arxiv.org/abs/nucl-th/0212090} {arXiv:nucl-th/0212090} \BibitemShut
  {NoStop}%
\bibitem [{\citenamefont {Xu}\ and\ \citenamefont {Ko}(2010)}]{Xu:2010ce}%
  \BibitemOpen
  \bibfield  {author} {\bibinfo {author} {\bibfnamefont {Jun}\ \bibnamefont
  {Xu}}\ and\ \bibinfo {author} {\bibfnamefont {Che~Ming}\ \bibnamefont {Ko}},\
  }\bibfield  {title} {\enquote {\bibinfo {title} {{Density matrix expansion
  for the MDI interaction}},}\ }\href {\doibase 10.1103/PhysRevC.82.044311}
  {\bibfield  {journal} {\bibinfo  {journal} {Phys. Rev. C}\ }\textbf {\bibinfo
  {volume} {82}},\ \bibinfo {pages} {044311} (\bibinfo {year} {2010})},\
  \Eprint {http://arxiv.org/abs/1007.1469} {arXiv:1007.1469 [nucl-th]}
  \BibitemShut {NoStop}%
\bibitem [{\citenamefont {Li}\ \emph {et~al.}(2018)\citenamefont {Li},
  \citenamefont {Cai}, \citenamefont {Chen},\ and\ \citenamefont
  {Xu}}]{Li:2018lpy}%
  \BibitemOpen
  \bibfield  {author} {\bibinfo {author} {\bibfnamefont {Bao-An}\ \bibnamefont
  {Li}}, \bibinfo {author} {\bibfnamefont {Bao-Jun}\ \bibnamefont {Cai}},
  \bibinfo {author} {\bibfnamefont {Lie-Wen}\ \bibnamefont {Chen}}, \ and\
  \bibinfo {author} {\bibfnamefont {Jun}\ \bibnamefont {Xu}},\ }\bibfield
  {title} {\enquote {\bibinfo {title} {{Nucleon Effective Masses in
  Neutron-Rich Matter}},}\ }\href {\doibase 10.1016/j.ppnp.2018.01.001}
  {\bibfield  {journal} {\bibinfo  {journal} {Prog. Part. Nucl. Phys.}\
  }\textbf {\bibinfo {volume} {99}},\ \bibinfo {pages} {29--119} (\bibinfo
  {year} {2018})},\ \Eprint {http://arxiv.org/abs/1801.01213} {arXiv:1801.01213
  [nucl-th]} \BibitemShut {NoStop}%
\bibitem [{\citenamefont {Oertel}\ \emph {et~al.}(2017)\citenamefont {Oertel},
  \citenamefont {Hempel}, \citenamefont {Kl\"ahn},\ and\ \citenamefont
  {Typel}}]{Oertel:2016bki}%
  \BibitemOpen
  \bibfield  {author} {\bibinfo {author} {\bibfnamefont {M.}~\bibnamefont
  {Oertel}}, \bibinfo {author} {\bibfnamefont {M.}~\bibnamefont {Hempel}},
  \bibinfo {author} {\bibfnamefont {T.}~\bibnamefont {Kl\"ahn}}, \ and\
  \bibinfo {author} {\bibfnamefont {S.}~\bibnamefont {Typel}},\ }\bibfield
  {title} {\enquote {\bibinfo {title} {{Equations of state for supernovae and
  compact stars}},}\ }\href {\doibase 10.1103/RevModPhys.89.015007} {\bibfield
  {journal} {\bibinfo  {journal} {Rev. Mod. Phys.}\ }\textbf {\bibinfo {volume}
  {89}},\ \bibinfo {pages} {015007} (\bibinfo {year} {2017})},\ \Eprint
  {http://arxiv.org/abs/1610.03361} {arXiv:1610.03361 [astro-ph.HE]}
  \BibitemShut {NoStop}%
\bibitem [{\citenamefont {Li}\ and\ \citenamefont {Han}(2013)}]{Li:2013ola}%
  \BibitemOpen
  \bibfield  {author} {\bibinfo {author} {\bibfnamefont {Bao-An}\ \bibnamefont
  {Li}}\ and\ \bibinfo {author} {\bibfnamefont {Xiao}\ \bibnamefont {Han}},\
  }\bibfield  {title} {\enquote {\bibinfo {title} {{Constraining the
  neutron-proton effective mass splitting using empirical constraints on the
  density dependence of nuclear symmetry energy around normal density}},}\
  }\href {\doibase 10.1016/j.physletb.2013.10.006} {\bibfield  {journal}
  {\bibinfo  {journal} {Phys. Lett. B}\ }\textbf {\bibinfo {volume} {727}},\
  \bibinfo {pages} {276--281} (\bibinfo {year} {2013})},\ \Eprint
  {http://arxiv.org/abs/1304.3368} {arXiv:1304.3368 [nucl-th]} \BibitemShut
  {NoStop}%
\bibitem [{\citenamefont {Lenk}\ and\ \citenamefont
  {Pandharipande}(1989)}]{Lenk:1989zz}%
  \BibitemOpen
  \bibfield  {author} {\bibinfo {author} {\bibfnamefont {R.~J.}\ \bibnamefont
  {Lenk}}\ and\ \bibinfo {author} {\bibfnamefont {V.~R.}\ \bibnamefont
  {Pandharipande}},\ }\bibfield  {title} {\enquote {\bibinfo {title} {{Nuclear
  mean field dynamics in the lattice Hamiltonian Vlasov method}},}\ }\href
  {\doibase 10.1103/PhysRevC.39.2242} {\bibfield  {journal} {\bibinfo
  {journal} {Phys. Rev. C}\ }\textbf {\bibinfo {volume} {39}},\ \bibinfo
  {pages} {2242--2249} (\bibinfo {year} {1989})}\BibitemShut {NoStop}%
\bibitem [{\citenamefont {Chen}\ \emph {et~al.}(2010)\citenamefont {Chen},
  \citenamefont {Ko}, \citenamefont {Li},\ and\ \citenamefont
  {Xu}}]{Chen:2010qx}%
  \BibitemOpen
  \bibfield  {author} {\bibinfo {author} {\bibfnamefont {Lie-Wen}\ \bibnamefont
  {Chen}}, \bibinfo {author} {\bibfnamefont {Che~Ming}\ \bibnamefont {Ko}},
  \bibinfo {author} {\bibfnamefont {Bao-An}\ \bibnamefont {Li}}, \ and\
  \bibinfo {author} {\bibfnamefont {Jun}\ \bibnamefont {Xu}},\ }\bibfield
  {title} {\enquote {\bibinfo {title} {{Density slope of the nuclear symmetry
  energy from the neutron skin thickness of heavy nuclei}},}\ }\href {\doibase
  10.1103/PhysRevC.82.024321} {\bibfield  {journal} {\bibinfo  {journal} {Phys.
  Rev. C}\ }\textbf {\bibinfo {volume} {82}},\ \bibinfo {pages} {024321}
  (\bibinfo {year} {2010})},\ \Eprint {http://arxiv.org/abs/1004.4672}
  {arXiv:1004.4672 [nucl-th]} \BibitemShut {NoStop}%
\bibitem [{\citenamefont {Li}\ and\ \citenamefont {Chen}(2005)}]{Li:2005jy}%
  \BibitemOpen
  \bibfield  {author} {\bibinfo {author} {\bibfnamefont {Bao-An}\ \bibnamefont
  {Li}}\ and\ \bibinfo {author} {\bibfnamefont {Lie-Wen}\ \bibnamefont
  {Chen}},\ }\bibfield  {title} {\enquote {\bibinfo {title} {{Nucleon-nucleon
  cross sections in neutron-rich matter and isospin transport in heavy-ion
  reactions at intermediate energies}},}\ }\href {\doibase
  10.1103/PhysRevC.72.064611} {\bibfield  {journal} {\bibinfo  {journal} {Phys.
  Rev. C}\ }\textbf {\bibinfo {volume} {72}},\ \bibinfo {pages} {064611}
  (\bibinfo {year} {2005})},\ \Eprint {http://arxiv.org/abs/nucl-th/0508024}
  {arXiv:nucl-th/0508024} \BibitemShut {NoStop}%
\bibitem [{\citenamefont {Pandharipande}\ and\ \citenamefont
  {Pieper}(1992)}]{Pandharipande:1992zz}%
  \BibitemOpen
  \bibfield  {author} {\bibinfo {author} {\bibfnamefont {V.~R.}\ \bibnamefont
  {Pandharipande}}\ and\ \bibinfo {author} {\bibfnamefont {Steven~C.}\
  \bibnamefont {Pieper}},\ }\bibfield  {title} {\enquote {\bibinfo {title}
  {{Nuclear transparency to intermediate-energy nucleons from (e, e'p)
  reactions}},}\ }\href {\doibase 10.1103/PhysRevC.45.791} {\bibfield
  {journal} {\bibinfo  {journal} {Phys. Rev. C}\ }\textbf {\bibinfo {volume}
  {45}},\ \bibinfo {pages} {791--798} (\bibinfo {year} {1992})}\BibitemShut
  {NoStop}%
\bibitem [{\citenamefont {Arndt}\ \emph {et~al.}(1977)\citenamefont {Arndt},
  \citenamefont {Hackman},\ and\ \citenamefont {Roper}}]{PhysRevC.15.1002}%
  \BibitemOpen
  \bibfield  {author} {\bibinfo {author} {\bibfnamefont {R.~A.}\ \bibnamefont
  {Arndt}}, \bibinfo {author} {\bibfnamefont {R.~H.}\ \bibnamefont {Hackman}},
  \ and\ \bibinfo {author} {\bibfnamefont {L.~D.}\ \bibnamefont {Roper}},\
  }\bibfield  {title} {\enquote {\bibinfo {title} {Nucleon-nucleon scattering
  analyses. ii. neutron-proton scattering from 0 to 425 mev and proton-proton
  scattering from 1 to 500 mev},}\ }\href {\doibase 10.1103/PhysRevC.15.1002}
  {\bibfield  {journal} {\bibinfo  {journal} {Phys. Rev. C}\ }\textbf {\bibinfo
  {volume} {15}},\ \bibinfo {pages} {1002--1020} (\bibinfo {year}
  {1977})}\BibitemShut {NoStop}%
\bibitem [{\citenamefont {Xia}\ \emph {et~al.}(2017)\citenamefont {Xia},
  \citenamefont {Xu}, \citenamefont {Li},\ and\ \citenamefont
  {Shen}}]{PhysRevC.96.044618}%
  \BibitemOpen
  \bibfield  {author} {\bibinfo {author} {\bibfnamefont {Yin}\ \bibnamefont
  {Xia}}, \bibinfo {author} {\bibfnamefont {Jun}\ \bibnamefont {Xu}}, \bibinfo
  {author} {\bibfnamefont {Bao-An}\ \bibnamefont {Li}}, \ and\ \bibinfo
  {author} {\bibfnamefont {Wen-Qing}\ \bibnamefont {Shen}},\ }\bibfield
  {title} {\enquote {\bibinfo {title} {Simulating spin dynamics with
  spin-dependent cross sections in heavy-ion collisions},}\ }\href {\doibase
  10.1103/PhysRevC.96.044618} {\bibfield  {journal} {\bibinfo  {journal} {Phys.
  Rev. C}\ }\textbf {\bibinfo {volume} {96}},\ \bibinfo {pages} {044618}
  (\bibinfo {year} {2017})}\BibitemShut {NoStop}%
\bibitem [{\citenamefont {Hua}\ and\ \citenamefont
  {Xu}(2024)}]{PhysRevC.109.034614}%
  \BibitemOpen
  \bibfield  {author} {\bibinfo {author} {\bibfnamefont {Lei-Ming}\
  \bibnamefont {Hua}}\ and\ \bibinfo {author} {\bibfnamefont {Jun}\
  \bibnamefont {Xu}},\ }\bibfield  {title} {\enquote {\bibinfo {title}
  {Transport properties of asymmetric nuclear matter in the spinodal region},}\
  }\href {\doibase 10.1103/PhysRevC.109.034614} {\bibfield  {journal} {\bibinfo
   {journal} {Phys. Rev. C}\ }\textbf {\bibinfo {volume} {109}},\ \bibinfo
  {pages} {034614} (\bibinfo {year} {2024})}\BibitemShut {NoStop}%
\bibitem [{\citenamefont {Xu}\ \emph {et~al.}(2024)\citenamefont {Xu} \emph
  {et~al.}}]{TMEP:2023ifw}%
  \BibitemOpen
  \bibfield  {author} {\bibinfo {author} {\bibfnamefont {Jun}\ \bibnamefont
  {Xu}} \emph {et~al.} (\bibinfo {collaboration} {TMEP}),\ }\bibfield  {title}
  {\enquote {\bibinfo {title} {{Comparing pion production in transport
  simulations of heavy-ion collisions at 270AMeV under controlled
  conditions}},}\ }\href {\doibase 10.1103/PhysRevC.109.044609} {\bibfield
  {journal} {\bibinfo  {journal} {Phys. Rev. C}\ }\textbf {\bibinfo {volume}
  {109}},\ \bibinfo {pages} {044609} (\bibinfo {year} {2024})},\ \Eprint
  {http://arxiv.org/abs/2308.05347} {arXiv:2308.05347 [nucl-th]} \BibitemShut
  {NoStop}%
\bibitem [{\citenamefont {Li}\ and\ \citenamefont {Ko}(1995)}]{Li:1995pra}%
  \BibitemOpen
  \bibfield  {author} {\bibinfo {author} {\bibfnamefont {Bao-An}\ \bibnamefont
  {Li}}\ and\ \bibinfo {author} {\bibfnamefont {Che~Ming}\ \bibnamefont {Ko}},\
  }\bibfield  {title} {\enquote {\bibinfo {title} {{Formation of superdense
  hadronic matter in high-energy heavy ion collisions}},}\ }\href {\doibase
  10.1103/PhysRevC.52.2037} {\bibfield  {journal} {\bibinfo  {journal} {Phys.
  Rev. C}\ }\textbf {\bibinfo {volume} {52}},\ \bibinfo {pages} {2037--2063}
  (\bibinfo {year} {1995})},\ \Eprint {http://arxiv.org/abs/nucl-th/9505016}
  {arXiv:nucl-th/9505016} \BibitemShut {NoStop}%
\bibitem [{\citenamefont {Chen}\ \emph {et~al.}(2003)\citenamefont {Chen},
  \citenamefont {Ko},\ and\ \citenamefont {Li}}]{Chen:2003ava}%
  \BibitemOpen
  \bibfield  {author} {\bibinfo {author} {\bibfnamefont {Lie-Wen}\ \bibnamefont
  {Chen}}, \bibinfo {author} {\bibfnamefont {C.~M.}\ \bibnamefont {Ko}}, \ and\
  \bibinfo {author} {\bibfnamefont {Bao-An}\ \bibnamefont {Li}},\ }\bibfield
  {title} {\enquote {\bibinfo {title} {{Light cluster production in
  intermediate-energy heavy ion collisions induced by neutron rich nuclei}},}\
  }\href {\doibase 10.1016/j.nuclphysa.2003.09.010} {\bibfield  {journal}
  {\bibinfo  {journal} {Nucl. Phys. A}\ }\textbf {\bibinfo {volume} {729}},\
  \bibinfo {pages} {809--834} (\bibinfo {year} {2003})},\ \Eprint
  {http://arxiv.org/abs/nucl-th/0306032} {arXiv:nucl-th/0306032} \BibitemShut
  {NoStop}%
\bibitem [{\citenamefont {Sun}\ and\ \citenamefont {Chen}(2017)}]{Sun:2017ooe}%
  \BibitemOpen
  \bibfield  {author} {\bibinfo {author} {\bibfnamefont {Kai-Jia}\ \bibnamefont
  {Sun}}\ and\ \bibinfo {author} {\bibfnamefont {Lie-Wen}\ \bibnamefont
  {Chen}},\ }\bibfield  {title} {\enquote {\bibinfo {title} {{Analytical
  coalescence formula for particle production in relativistic heavy-ion
  collisions}},}\ }\href {\doibase 10.1103/PhysRevC.95.044905} {\bibfield
  {journal} {\bibinfo  {journal} {Phys. Rev. C}\ }\textbf {\bibinfo {volume}
  {95}},\ \bibinfo {pages} {044905} (\bibinfo {year} {2017})},\ \Eprint
  {http://arxiv.org/abs/1701.01935} {arXiv:1701.01935 [nucl-th]} \BibitemShut
  {NoStop}%
\bibitem [{\citenamefont {Charity}\ \emph {et~al.}(1988)\citenamefont {Charity}
  \emph {et~al.}}]{Charity:1988zz}%
  \BibitemOpen
  \bibfield  {author} {\bibinfo {author} {\bibfnamefont {R.~J.}\ \bibnamefont
  {Charity}} \emph {et~al.},\ }\bibfield  {title} {\enquote {\bibinfo {title}
  {{Systematics of complex fragment emission in niobium-induced reactions}},}\
  }\href {\doibase 10.1016/0375-9474(88)90542-8} {\bibfield  {journal}
  {\bibinfo  {journal} {Nucl. Phys. A}\ }\textbf {\bibinfo {volume} {483}},\
  \bibinfo {pages} {371--405} (\bibinfo {year} {1988})}\BibitemShut {NoStop}%
\bibitem [{\citenamefont {Charity}(2010)}]{Charity:2010wk}%
  \BibitemOpen
  \bibfield  {author} {\bibinfo {author} {\bibfnamefont {R.~J.}\ \bibnamefont
  {Charity}},\ }\bibfield  {title} {\enquote {\bibinfo {title} {{A Systematic
  description of evaporation spectra for light and heavy compound nuclei}},}\
  }\href {\doibase 10.1103/PhysRevC.82.014610} {\bibfield  {journal} {\bibinfo
  {journal} {Phys. Rev. C}\ }\textbf {\bibinfo {volume} {82}},\ \bibinfo
  {pages} {014610} (\bibinfo {year} {2010})},\ \Eprint
  {http://arxiv.org/abs/1006.5018} {arXiv:1006.5018 [nucl-th]} \BibitemShut
  {NoStop}%
\bibitem [{\citenamefont {Wang}\ \emph {et~al.}(2021)\citenamefont {Wang},
  \citenamefont {Huang}, \citenamefont {Kondev}, \citenamefont {Audi},\ and\
  \citenamefont {Naimi}}]{Wang:2021xhn}%
  \BibitemOpen
  \bibfield  {author} {\bibinfo {author} {\bibfnamefont {Meng}\ \bibnamefont
  {Wang}}, \bibinfo {author} {\bibfnamefont {W.~J.}\ \bibnamefont {Huang}},
  \bibinfo {author} {\bibfnamefont {F.~G.}\ \bibnamefont {Kondev}}, \bibinfo
  {author} {\bibfnamefont {G.}~\bibnamefont {Audi}}, \ and\ \bibinfo {author}
  {\bibfnamefont {S.}~\bibnamefont {Naimi}},\ }\bibfield  {title} {\enquote
  {\bibinfo {title} {{The AME 2020 atomic mass evaluation (II). Tables, graphs
  and references}},}\ }\href {\doibase 10.1088/1674-1137/abddaf} {\bibfield
  {journal} {\bibinfo  {journal} {Chin. Phys. C}\ }\textbf {\bibinfo {volume}
  {45}},\ \bibinfo {pages} {030003} (\bibinfo {year} {2021})}\BibitemShut
  {NoStop}%
\end{thebibliography}%
\end{document}